\documentclass[a4paper,3p]{elsarticle}
\usepackage{amsmath,amsfonts,amssymb}
\usepackage{graphicx}
\usepackage{xcolor}

\usepackage{algpseudocode}
\usepackage[ruled]{algorithm}

\usepackage{caption}
\usepackage{subcaption}
\usepackage{hyperref}

\DeclareMathOperator{\tr}{tr}
\renewcommand\Re{{\rm Re}}
\newcommand\We{{\rm We}}
\newcommand\Ar{{\rm Ar}}
\newcommand\Bo{{\rm Bo}}
\newcommand\Ca{{\rm Ca}}
\begin{document}
\title{Manifold death: a Volume of Fluid implementation of controlled topological changes in thin sheets by the signature method}

\author[IJLRA]{Leonardo Chirco\corref{cor}}
\ead{leonardo.chirco@sorbonne-universite.fr}

\author[IJLRA]{Jacob Maarek}
\ead{jacob.maarek@sorbonne-universite.fr}

\author[IJLRA]{St\'ephane Popinet}
\ead{popinet@basilisk.fr}

\author[IJLRA,IUF]{St\'ephane Zaleski}
\ead{stephane.zaleski@sorbonne-universite.fr}

\address[IJLRA]{Sorbonne Universit\'e and CNRS, Institut Jean Le Rond d'Alembert 
UMR 7190, F-75005 Paris, France}
\address[IUF]{Institut Universitaire de France, Paris, France}

\cortext[cor]{Corresponding author}

\begin{abstract}
A well-known drawback of the Volume-Of-Fluid (VOF) method is that
the breakup of thin liquid films or filaments is mainly caused 
by numerical aspects rather than by physical ones. 
The rupture of thin films occurs when their thickness reaches 
the order of the grid size and by refining the grid the breakup events are delayed.
When thin filaments rupture, many droplets are generated 
due to the mass conserving properties of VOF. 
Thus, the numerical character of the breakup does not allow 
obtaining the desired convergence of the droplet size distribution upon grid refinement.
In this work, we present a novel algorithm to detect and perforate thin structures.
First, thin films or ligaments are identified 
by taking quadratic moments of an indicator obtained from the volume fraction. 
A multiscale approach allows us to choose the critical film thickness independently of the mesh resolution.
Then, the breakup is induced by making holes in the films
before their thickness reaches the grid size. 
We show that the method improves the convergence upon grid refinement
of the droplets size distribution and of enstrophy. 
\end{abstract}

\maketitle

\section{Introduction}
Multiphase fluid mechanics with sharp interfaces 
involves diverse singularities each with its own difficulties.
In this paper we address the change of topology happening 
when a hole forms in a thin liquid sheet 
and in particular its numerical modelling in Volume-Of-Fluid (VOF) methods.
This topology change has important consequences for fluid fragmentation,
and in particular for the droplet size distribution.
Several mechanisms for the rupture of thin sheets have been proposed \cite{villermaux2020fragmentation,lohse2020double},
but a full description is still missing.
In some cases, inter-molecular forces may be shown
to have a destabilizing effect on thin sheets,
while in other cases, as in some soap films,
permanent dipoles may form on interfaces having a stabilizing effect \cite{Israelachvili}. In this paper we consider interfaces 
that do not display such a stabilizing effect 
and break before reaching molecular length scales.

Mathematically, the Navier-Stokes equations 
with the assumption of sharp interfaces between phases 
are incomplete as they do not describe the piercing of thin sheets.
These equations can only describe the indefinite thinning of a sheet
when it is compressed or stretched by the flow.
Eventually, it may happen that the sheet becomes thinner than molecular scales,
at which point the equations cease to be valid.
However, before this happens, molecular forces may induce the breakup of films
\cite{vrij1966possible,ruckenstein1974spontaneous,
radoev1983critical,zhang1999similarity,moreno2020stokes}. 
In many cases, however, and in particular in atomization 
or floating bubble experiments,
observations show that sheets are pierced
well before their thickness reaches molecular length scales
\cite{Opfer14,poulain2018ageing}.
Thus, an ad-hoc prescription for piercing thin sheets of
some macroscopic length scale is needed to describe real flows. 

In addition to the physical modelling challenge, 
the numerical simulation of thin sheets and their breakup
poses significant challenges.
In fact, when the fluid structures become too thin 
they can not be correctly resolved
with the Volume-of-Fluid (VOF) method. 
The most popular interface reconstruction approaches are responsible
for the breakup of thin sheets
whose thickness is of the order of the grid size
and for the formation of several droplets.
To avoid the occurrence of artificial breakup using VOF,
extremely fine grids must be used,
such that $\Delta<h$ 
(where $\Delta$ is the cell size and $h$ the film thickness) 
and huge computational efforts are required even
using codes with adaptive mesh refinement (AMR) capabilities
\cite{ling2017spray,agbaglah2021breakup}.
Otherwise, as soon as the thickness of the liquid sheet 
approaches the cell size ($\Delta\approx h$), 
the artificial breakup of thin sheets
does not permit to observe convergence upon grid refinement
of some meaningful quantities such as the droplets size distribution or enstrophy.
The standard Level-Set (LS) method \cite{sethian1999level}
does not guarantee mass conservation 
and mass/volume evaporation occurs when the filaments
become thinner than the cell size.
Thus, various techniques, such as particle LS \cite{zhao2018one,wang2022local}
or mass conserving LS \cite{yuan2018simple}
have been developed to reduce this limitation.
Anyhow, an intrinsic breakup length scale (equal to the grid size or smaller,
depending on the LS method) exists and the rupture of thin structures with
high curvature is still dependent on the grid size.
Finally, Front-Tracking (FT) \cite{trygg2001} methods do not include
automatic topology modification and breakup/reconnection algorithms must be
implemented.

Some methods have been proposed to detect thin ligaments or sheets.
In \cite{jemison2015filament} a Multimaterial Moment-of-Fluid method
is used to capture under-resolved filaments thinner than the grid cells,
while a two-plane reconstruction to track sub-grid scale sheets
has been recently proposed in \cite{chiodi2020adva}.
A geometric approach is used in \cite{henri2022geom} to detect 
the medial axis in a Level-Set framework.
Automatic reconnection algorithms have been suggested for thin sheets in the
Front-Tracking method \cite{lu2018direct}.
A limitation of the work is that the authors decide to use a critical
reconnection thickness proportional to the cell size
and therefore the breakup process remains grid-dependent.
Some algorithms have also been suggested to make ``numerical''
reconnection impossible for the VOF method  \cite{focke2013collision}.
We are not aware of any similar mechanisms 
(either promoting reconnection or blocking it) for the Level-Set method. 

In this paper we propose a \textit{manifold death} algorithm 
to perforate sheets or ligaments of a given thickness 
in a controlled way, thus avoiding the numerical breakup that affects the VOF method and the associated grid convergence issues. 
First, the thin regions are detected by computing quadratic forms
based on the values of the color function. 
The signs of the eigenvalues of the quadratic form,
known as \textit{signature}, are then used to identify thin regions. 
Once their position is known, we want to create artificial holes 
that expand and ``destroy'' the thin regions.
This leads to several questions, such as: 
how many holes are needed to perforate a thin sheet,
where are they placed, when do they appear?
However, these are still open questions.
In fact, many complex phenomena are involved in
the physics of hole nucleation that
leads to the rupture of thin sheets, see \cite{lohse2020double}.
For example, in the experiments of \cite{kant2022bags} 
one or several holes are seen
but their origin is still not known nor
is the effect of the number of holes investigated.
With this in mind, when developing the manifold death method,
to answer the prior questions we tried to maintain a balance between 
the ambition to replicate the observations and a feasible numerical implementation.
This work has to be intended as a preliminary study, in particular about
the hole formation aspects and parameters,
and further improvements and comparisons will be performed in the future.

The rest of the paper is organized as follows. 
The mathematical model is defined in Section \ref{sec:eq}.  
The manifold death algorithm is described in Section \ref{sec:manif}.
The numerical results obtained 
with the manifold death algorithm are discussed in Section \ref{sec:resu},
followed by our conclusions.

\section{Governing equations} \label{sec:eq}
The governing equations for the incompressible two-phase flow
with immiscible fluids are written using the one-fluid formulation, see \cite{Tryggvason}.
A Heaviside function $H$ is used to locate the interface,
such that $H=1$ in fluid 1 and $H=0$ elsewhere.
To solve numerically the problem, the 
discontinuous $H$ function is replaced with the discretized volume fraction (or color function)
$C$ defined as the average value of $H$ in each computational cell.
The volume fraction $C$ is then a piece-wise constant approximation of $H$
such that $C\in[0,1]$, where 
fractional values are associated to the cells cut by the interface.
Then, in the one-fluid formulation, 
the value of a generic material property $\pi$ is 
a function of $C$, namely $\pi=C\pi_1+(1-C)\pi_2$.
The volume fraction is obtained by solving
the following advection equation
\begin{equation}
\frac{\partial C}{\partial t}+ {\bf u} \cdot \nabla C = 0 \,.
\label{eq:vof}
\end{equation}
The velocity ${\bf u}$ and pressure $p$ are obtained by solving 
the Navier-Stokes equations that express the conservation of mass and momentum
\begin{align}
\frac{\partial\rho {\bf u}}{\partial t}+
\nabla\cdot(\rho{\bf u}{\bf u})&=
-\nabla p+\nabla \cdot 
\left[\mu\left(\nabla{\bf u}+\nabla {\bf u}^T\right)\right]+
\rho {\bf g}+ {\bf f} \label{eq:mom}\,, \\
\nabla \cdot {\bf u}&=0 \label{eq:div}\,,
\end{align}
where $\rho$ and $\mu$ are the density and viscosity, respectively.
The gravitational force is taken into account with the $\rho \mathbf{g}$ term.
Surface tension is modeled with the term 
$\mathbf{f}= \sigma\kappa \mathbf{n}\delta_S$,
where $\sigma$ is the surface tension, $\kappa$
denotes the curvature of the interface,
$\mathbf{n}$ is the unit vector normal to the interface 
and $\delta_S$ is the surface Dirac distribution on the interface
(i.e. zero everywhere except on the interface).

The open-source code Basilisk is used to solve
the two-phase incompressible Navier-Stokes system with
quad/octrees \cite{popinet2015quadtree}.
An efficient adaptive technique based on a wavelet decomposition 
of the volume fraction and velocity fields
allows solving with a high resolution
only in the relevant parts of the domain, thus reducing
the computational cost of the simulation \cite{van2018towards}.
A momentum-conserving scheme is used for the velocity.
A piece-wise linear geometric VOF method
is adopted for tracking the interface and
a well-balanced Continuous Surface Force (CSF) method coupled with height functions
is used to calculate the surface tension term
\cite{popinet2018numerical}.

\section{The manifold death algorithm} \label{sec:manif}
In this section we describe the manifold death method 
used to perforate thin structures in a controlled way.
\subsection{The signature method}
\begin{figure}[htb!]
 \centering
 \includegraphics[width=0.7\textwidth]{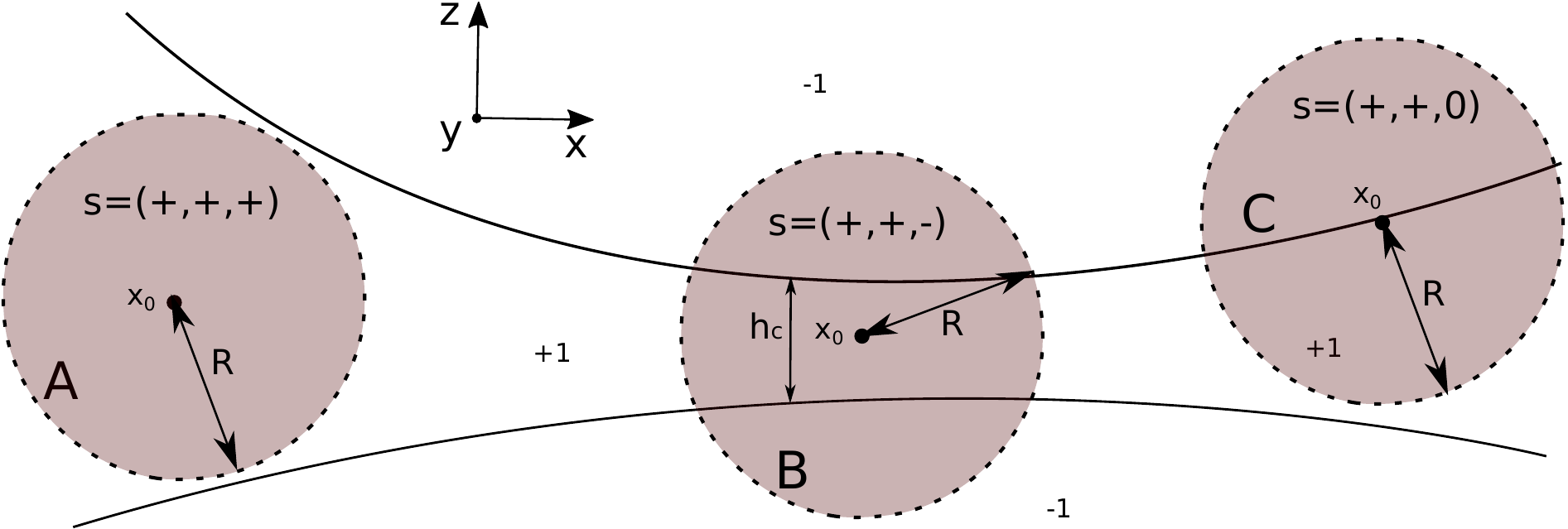}
  \caption{Graphical representation of the signature method applied 
  to a sheet infinite in the $y$ direction. 
   A: bulk of phase. B: thin sheet. C: interface.}
  \label{imm:signature}
\end{figure}
\alglanguage{pseudocode}
\begin{algorithm}[!htb]
\begin{algorithmic}[0]
 \State  1. Set $\phi=2C-1$. \Comment{\emph{Symmetric indicator function}}
 \ForAll {cells}
    \State 2. Translate the coordinates as $x_i\gets x_i-x_{cell}$ with $i=1,\dots,3$.
    \State 3. Compute the quadratic moments $T_{ij}$ using Eq. \eqref{eq:quadMom}.
    \State 4. Compute the eigenvalues $\epsilon_i$ of $\mathbf{T}$ to build the signature $s$.
    \State 5. Locate the cell using Table \ref{tab:sign}.
    \EndFor
\end{algorithmic}
\caption{\label{alg:sign} The signature method}
\end{algorithm}
\begin{table}[htb!]
\centering
   \begin{tabular}{ccc} 
     \hline\hline
      \multicolumn{2}{c}{Signature $s$} & Position of $\mathbf{x}_0$\\
      3D &2D&                   \\
      \hline
      $(+,+,+)$&$(+,+)$ & Bulk of the phase  \\
      $(+,+,-)$&/   & Sheet              \\
      $(+,-,-)$&$(+,-)$ & Ligament           \\
      $(+,+,0)$&$(+,0)$ & Interface          \\
      $(+,-,0)$&/    & Edge of a sheet    \\
      $(-,-,0)$&$(-,0)$ & Edge of a ligament \\
     \hline\hline
  \end{tabular}
  \caption{Examples of the signature in three and two dimensions,
  see Figure \ref{imm:signature}.}
    \label{tab:sign}
\end{table}
The first step of the manifold death algorithm consists in detecting
the thin sheets (or ligaments) in the domain. 
To do that, we use the following \textit{signature method}:
\begin{enumerate}
\item Consider a point $\mathbf{x}_0 \in \mathbb{R}^3$ 
      and translate the coordinate system so that    
      the new origin is $\mathbf{x}_0 = \mathbf{0}$. Consider a
      radius $R$ approximately the same size of the sheet thickness $h_c$
      one wants to detect, see Figure \ref{imm:signature}, 
      and the bilinear form
      $f (\mathbf{x}, \mathbf{x}) = x_i x_j T_{ij}$.
\item The quadratic moments $T_{ij}$ on a sphere $S$ of radius $R$
      can be found by integrating
      \begin{equation} \label{eq:quadMom}
        T_{ij} = \int_V x'_i x'_j \phi(\mathbf{x}') \rm{d}\mathbf{x}'   \,,
      \end{equation}
      where $\phi=2C-1$ is a symmetric indicator function.
\item After orthonormalization of the quadratic form, one finds
      a new set of coordinates in which
      $f (\mathbf{X}, \mathbf{X}) = \epsilon_i X_i^2 $, 
      where $\epsilon_i$ are the eigenvalues of the operator with matrix $\mathbf{T}$.
      The number of positive, negative and zero values of $\epsilon_i$
      is the signature $s$ of the quadratic form.
      For the given point $\mathbf{x}_0$, the signature $s$ indicates the average shape
      of the interface in the vicinity of the point 
      and can be used to determine whether the point is in the bulk of the phase,
      in a thin sheet (or ligament) or close to the interface, 
      see Figure \ref{imm:signature} and Table \ref{tab:sign}. 
      
\end{enumerate}
The eigenvalues of $\mathbf{T}$ can be found by solving the equation
\begin{equation}
 \det(\mathbf{T} - \epsilon \mathbf{I}) = 0 \,,
\end{equation}
that in two dimensions gives
$\epsilon_{1,2}=\tr(\mathbf{T})\pm\sqrt{(\tr(\mathbf{T})^2-4\Delta)}/2$.
For computing the eigenvalues, we rely on the GNU Scientific Library \cite{galassi2002gnu}. 
Table \ref{tab:sign} lists some of the signatures
that can be obtained using this method, while
a concise description of the signature method is given in Algorithm \ref{alg:sign}.

To clarify the rationale for the signatures
reported in Table \ref{tab:sign}, let us introduce the following example.
Consider a sheet aligned with the axis, infinite in the $y$ direction
and of thickness $h_c$ in the $z$ direction, see Figure \ref{imm:signature}.
We first recall that the signature of a matrix is invariant upon rotation of the coordinate system.
From symmetry, the non-diagonal terms of $\mathbf{T}$ cancel out or are negligible
for a slightly deformed object and only the diagonal terms remain:
$T_{xx},T_{yy},T_{zz}=\int x_i^{\prime 2}\phi(\mathbf{x'})\mathrm{d}\mathbf{x'}$,
which means that we are integrating along the Cartesian axis.
These remaining diagonal terms are the eigenvalues of $\mathbf{T}$
and their signs are the signature we are interested in. 
When we evaluate $T_{xx}=\int x^{\prime 2}\phi(\mathbf{x'})\mathrm{d}\mathbf{x'}$,
since we are inside phase 1 we have $C=1$ and $\phi=1$ 
so that $T_{xx}$ is positive. 
The same arguments can be used to show that also $T_{yy}=\int y^{\prime 2}\phi(\mathbf{x'})\mathrm{d}\mathbf{x'}$ is positive.
When we consider $T_{zz}=\int z^{\prime 2}\phi(\mathbf{x'})\mathrm{d}\mathbf{x'}$, if we use for the integration a sphere of radius $R<h_c/2$ 
(sphere A in Figure \ref{imm:signature}),
we are again inside the reference phase, $T_{zz}=1$,
and we recover the signature $(+,+,+)$
typical of the bulk of phase.
If instead we take $R>h_c/2$ (sphere B in the same figure), 
we are no longer in the reference phase so $C=0$, $\phi=-1$,
$T_{zz}$ is negative and the signature $(+,+,-)$ identifies a thin sheet.
For a ligament (consider doing a revolution around the horizontal axis of
Figure \ref{imm:signature}) aligned with one of the axis (e.g. the $x$ one),
the signature becomes $(+,-,-)$,
since the integration of $T_{yy}$ now yields a negative value.
Finally, zero values are obtained when the center of the sphere used for
the integration lies on the interface (sphere C in the same figure),
since the positive and negative contributions cancel out. 
Then the signature $(+,-,0)$ denotes the edge of a sheet,
while $(-,-,0)$ the edge of a ligament. 
From a numerical point of view, we use a heuristically
chosen threshold (0.1) to detect zero eigenvalues.
\begin{figure}[tb!]
 \centering
  \includegraphics[width=.4\textwidth]{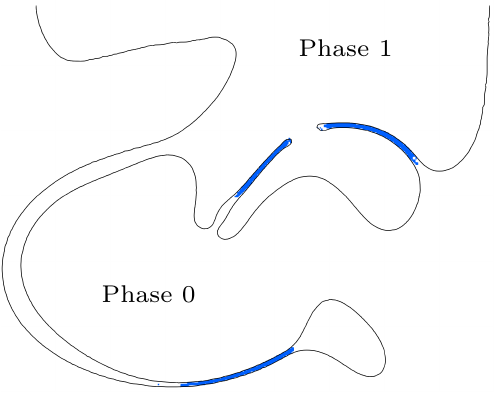}
  \caption{Slice of a 3D simulation. Interface in black and thin sheets in blue.}
  \label{imm:sig}
\end{figure}
The thin sheets identified with the signature method are
shown in Figure \ref{imm:sig}. The figure clearly shows that
the method is symmetric. 
In fact, the algorithm detects a thin sheet in the reference phase in the bottom,
while, in the two other regions above,
the reference phase 1 encloses thin sheets of the phase 0.
To summarize, with this method we can identify thin regions in both phases.

\subsection{Basilisk implementation}
\begin{figure}[tb!]
 \centering
  \includegraphics[width=.7\textwidth]{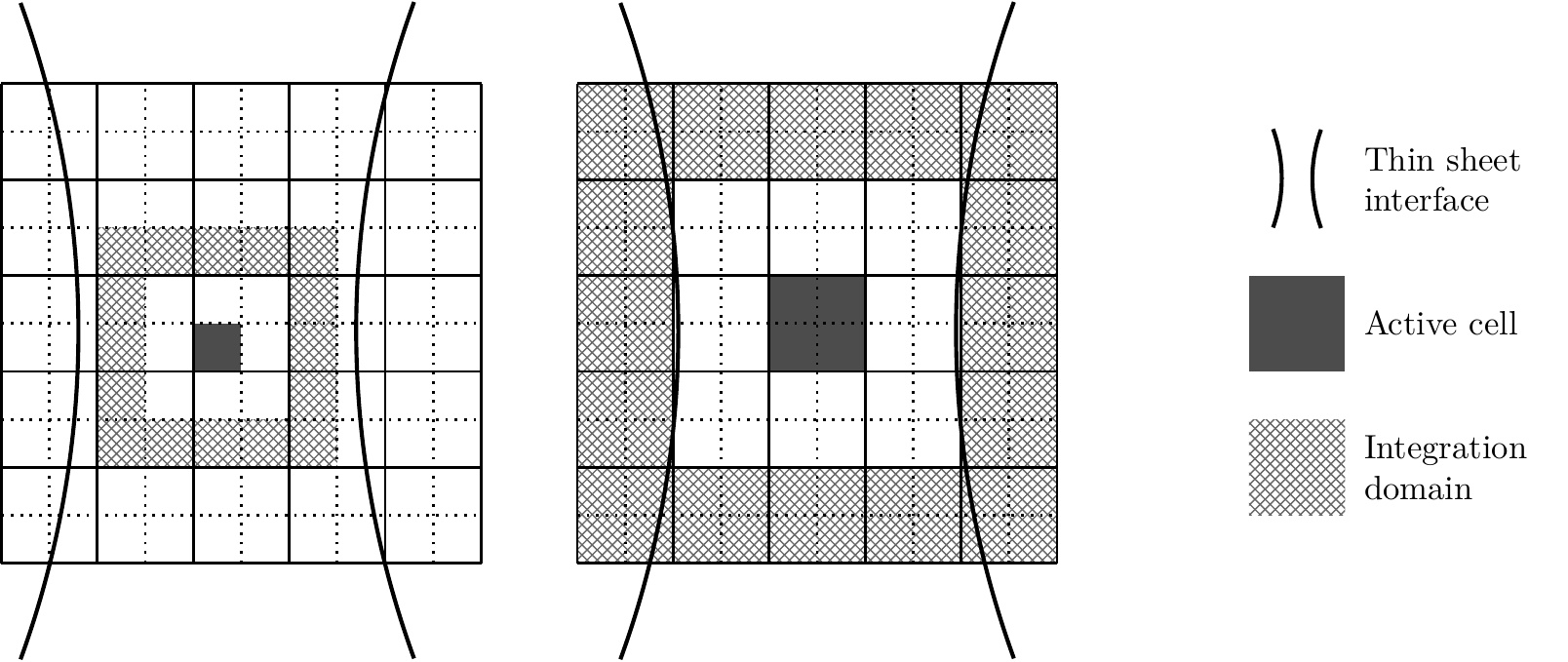}
  \caption{Multilevel implementation of the signature method. 
  Domains used for the numerical integration of \eqref{eq:quadMom} (hatched cells)
  on a fine (left) and coarse (right) grid around the shaded active cell $\mathbf{x}_0$.
  The detectable critical thickness $h_c$ on the coarse grid (right)
  is twice the one on the finer grid (left). }
  \label{imm:stencil}
\end{figure}

For the numerical computation of the quadratic moments
in \eqref{eq:quadMom}, some simplifications 
and adjustments are necessary to tailor the method to the Basilisk code.
For simplicity, we first replace the sphere 
with a cube centered in $\mathbf{x}_0$ of size $L=2R$. 
The numerical integration of \eqref{eq:quadMom} is done on a shell 
of thickness $\Delta$, the hatched cells in Figure \ref{imm:stencil}.
Moreover, in order to take advantage of Basilisk's capabilities,
we make sure that the cube lies completely within
the stencil (i.e. the set of surrounding cells) around $\mathbf{x}_0$.
Since the standard stencil size in Basilisk is five cells ($5\Delta$),
we can detect thin sheets whose thickness is approximately three cells ($3\Delta$).
To identify larger sheets ($h_c>3\Delta$) we take advantage 
of the multilevel nature of Basilisk
by computing the signature on the appropriate coarser grid.
In fact, after the first step of Algorithm \ref{alg:sign}, 
we restrict the values of
$\phi$ to the appropriate coarser grid and, at the end of the algorithm, 
we prolongate back the obtained signatures onto finer grids.
The restriction operator used for the indicator function $\phi$
sets to the coarse \textit{parent} cells the
average value of the four (eight in 3D) corresponding \textit{children} cells.
The prolongation operator for the signature simply attributes
to the children cells
the value of their parent cell.
Therefore, a critical thin sheet thickness
can be set independently from the mesh resolution.
Figure \ref{imm:stencil} illustrates the multilevel approach
used for detecting thin structures.
Using the finer grid (left), the distance between the two interfaces
is larger than $h_c \approx 3\Delta_{fine}$ 
and the shaded cell is marked as ``bulk of phase'',
while on the coarser one a thin structure
of thickness $h_c \approx 3\Delta_{coarse}=6\Delta_{fine}$ is detectable.

We would like to remark that the algorithm is compatible
with the adaptive-grid
approach of Basilisk and then it is suitable for large parallel computation.
Also, our observations indicate that the computational time
required for computing the signature is small compared to the time used by
the multiphase solver to advance the solution.
In Table \ref{tab:cpu} we give the CPU time for ten time steps
of the multiphase solver with and without the signature method.
For these measurements, we take our last numerical test  
(phase inversion, Section \ref{sec:phaseInv}) using a $512^3$ maximum resolution 
and the signature is computed with three different $h_c$.
The computational overhead introduced by the signature method ranges
from 1.9\% to 7.2\% of the reference time. 
By considering larger $h_c$ the overhead reduces, 
since the number of active cells becomes smaller.
This is due to the use of the multilevel approach on adaptive octree grids,
in fact, with every restriction operation, groups of eight children cells
are replaced by one common parent cell.
Moreover, the full manifold death algorithm 
(of which the signature method is the first step)
should not be invoked at every time step (see Section \ref{sec:topo}) and then
the computational overhead attributable to the signature method
can be considered negligible.

Finally, the signature method can be used also as a criterion
to enforce adaptive grid refinement in thin regions, see \cite{zhang2021three}. 
\begin{table}[htb!]
\centering
   \begin{tabular}{cccc} 
     \hline\hline
     Test case & CPU time (s)& Relative CPU  & Nb active cells \\
     \hline
     Multiphase solver       & 74.06  &1     & $1\,825\,064$   \\
     $h_{c1}=3\Delta_{128}$  & 75.46  &1.019 & $406\,840$      \\
     $h_{c2}=3\Delta_{256}$  & 77.26  &1.043 & $1\,180\,664$   \\
     $h_{c3}=3\Delta_{512}$  & 79.40  &1.072 & $1\,825\,064$   \\
     \hline\hline
  \end{tabular}
  \caption{CPU time and number of active cells for the reference
  case without signature method and with the signature method
  with three values of thin sheet thickness $h_c$. 
  $\Delta_{N}$ is equal to $L/N$, where $L$ is the domain size and $N$
  the number of cells per side.}
  \label{tab:cpu}
\end{table}

\subsection{The topology changes}\label{sec:topo}
\alglanguage{pseudocode}
\begin{algorithm}[!htb]
\begin{algorithmic}[0]
 \Require The signature $\epsilon$ for every cell.
 \ForAll {cells belonging to thin regions}
        \State 1. Randomly choose the position of the holes. 
        \Comment{\emph{Create at most $n_h$ holes}}
        \If {cell to perforate}
            \State 2. Compute average $C$ in the stencil around cell.
            \If {$C<C^*$}
                \State 3. $C=1$. \Comment{\emph{Holes made in phase 0}}
                \Else 
                \State 4. $C=0$. \Comment{\emph{Holes made in phase 1}}
        \EndIf
    \EndIf
 \EndFor
 \State 5. Advance the solution by $t_h$ before calling again this algorithm.
\end{algorithmic}
\caption{\label{alg:holes} The creation of holes}
\end{algorithm}
Once we have identified the thin sheets with the signature method,
we want to perforate them by creating holes
before their thickness reaches the cell size and the numerical breakup happens.
Since a universal theory for the film breaking mechanism
has not yet been developed (see \cite{lohse2020double}),
we decide to select randomly the cells 
(among those in thin sheets) where a hole is created.

For this part of the manifold death method we need to introduce many
parameters whose values have been chosen empirically.
More sophisticated versions and a broader investigation 
of the effect of these parameters on the method
will be presented in future works.
When the previously detected thin regions are in the reference phase
(i.e. $C>C^*$), we change the topology by placing some cubic holes
setting the color function $C$ to zero. 
Otherwise, to force interface reconnections 
when the thin region is in the other phase ($C<C^*$),
we set the color function $C$ to 1. 
In our simulations we take $C^*=0.3$.

An important remark is that 
the size of the holes (made in both phases) is critical, in fact holes will expand 
only if their diameter is at least the sheet thickness $h_c$,
otherwise they will re-close to minimize the energy of the system
\cite{taylor1973making}.
In order to recover the desired grid independency, 
in addition to the thickness $h_c$, also the size of the holes 
has to be fixed and independent of the cell size. 
To this aim, we create cubic holes of the same length $h_c$
used for the computation of the signature.
By choosing the minimum size for which the expansion criterion is satisfied,
we are minimizing the perturbations on the system 
as well as the amount of mass removed or added.
Similarly, we want to minimize the number of holes that we create,
since they have an impact on the overall dynamics of the retracting rim.
Thus, the manifold death algorithm is used only at fixed time interval $t_h$
to give the holes some time to expand with Taylor-Culick velocity
and we limit the maximum number of holes $n_h$ created at every iteration.
In the limit where the manifold death is used at every iteration
of the multiphase solver, which means that $t_h$ is equal to the presumably small
time step, the holes do not have enough time to expand
and the thin sheets gets replaced
by a large number of holes in an unphysical manner.
On the contrary, if $t_h$ is too large the unwanted numerical breakup happens.
A description of the method is reported in Algorithm \ref{alg:holes}.

\section{Results} \label{sec:resu}
In this section we present our results. 
We apply the manifold death method to a simple single vortex test,
to an axisymmetric secondary atomization situation and to a phase inversion problem.
\subsection{Single vortex}
\begin{figure}[htb!]
 \centering
 \begin{subfigure}[t]{0.45\textwidth}
  \includegraphics[width=\textwidth]{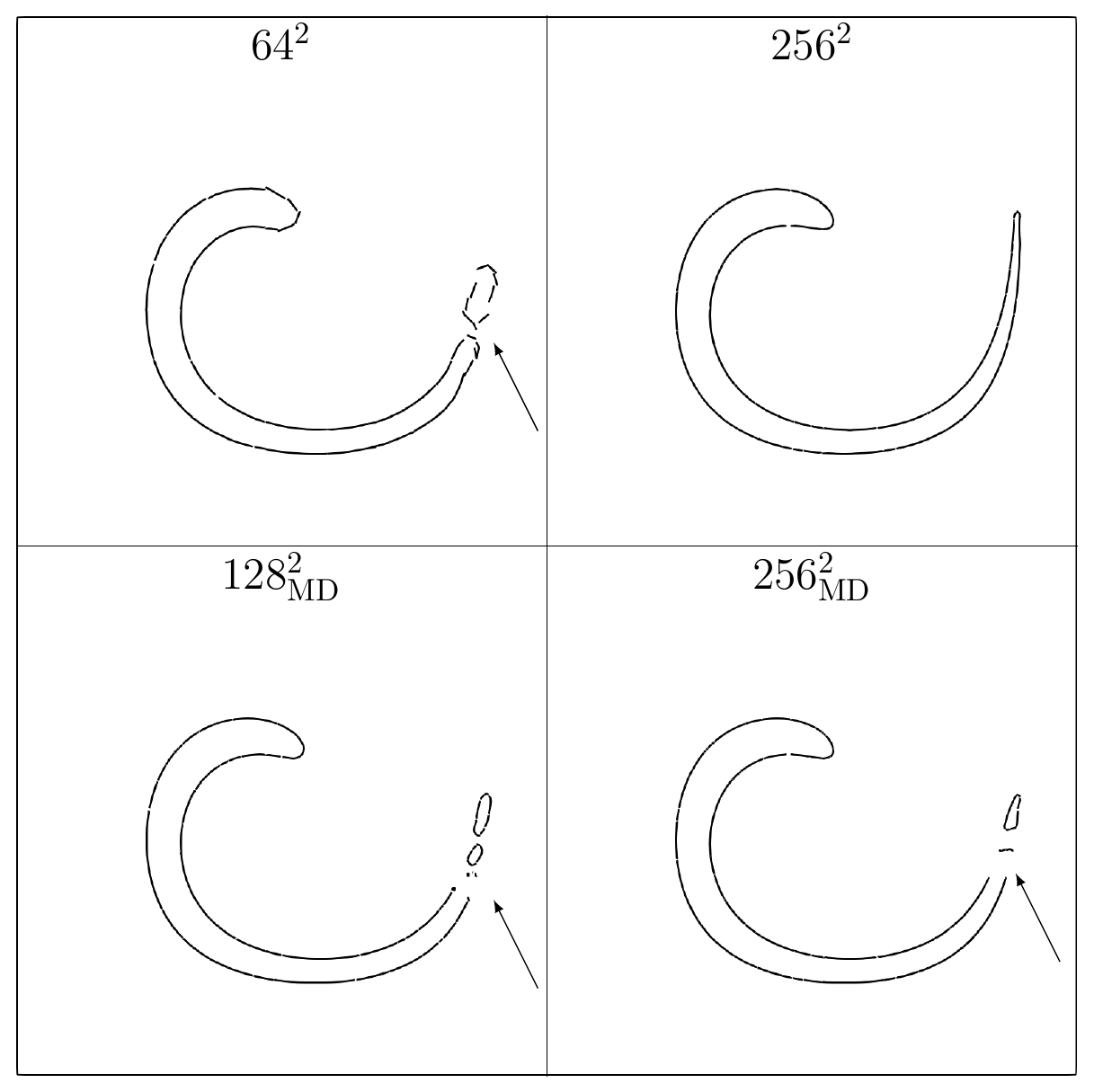}
  \caption{$t=1.1$ s.}
 \end{subfigure}
 \begin{subfigure}[t]{0.45\textwidth}
  \includegraphics[width=\textwidth]{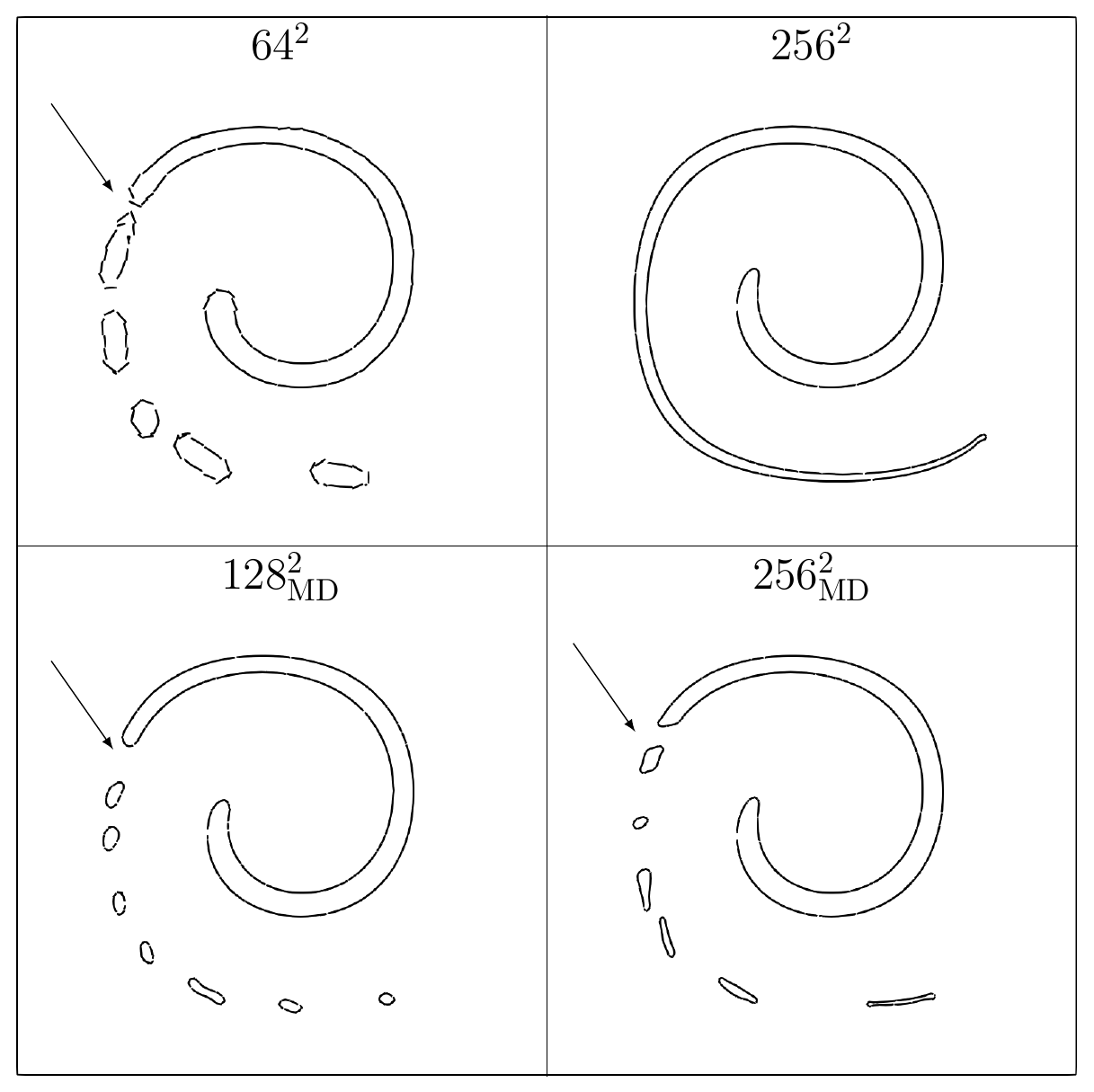}
  \caption{$t=1.9$ s.}  
  \end{subfigure}
  \begin{subfigure}[b]{0.45\textwidth}
  \includegraphics[width=\textwidth]{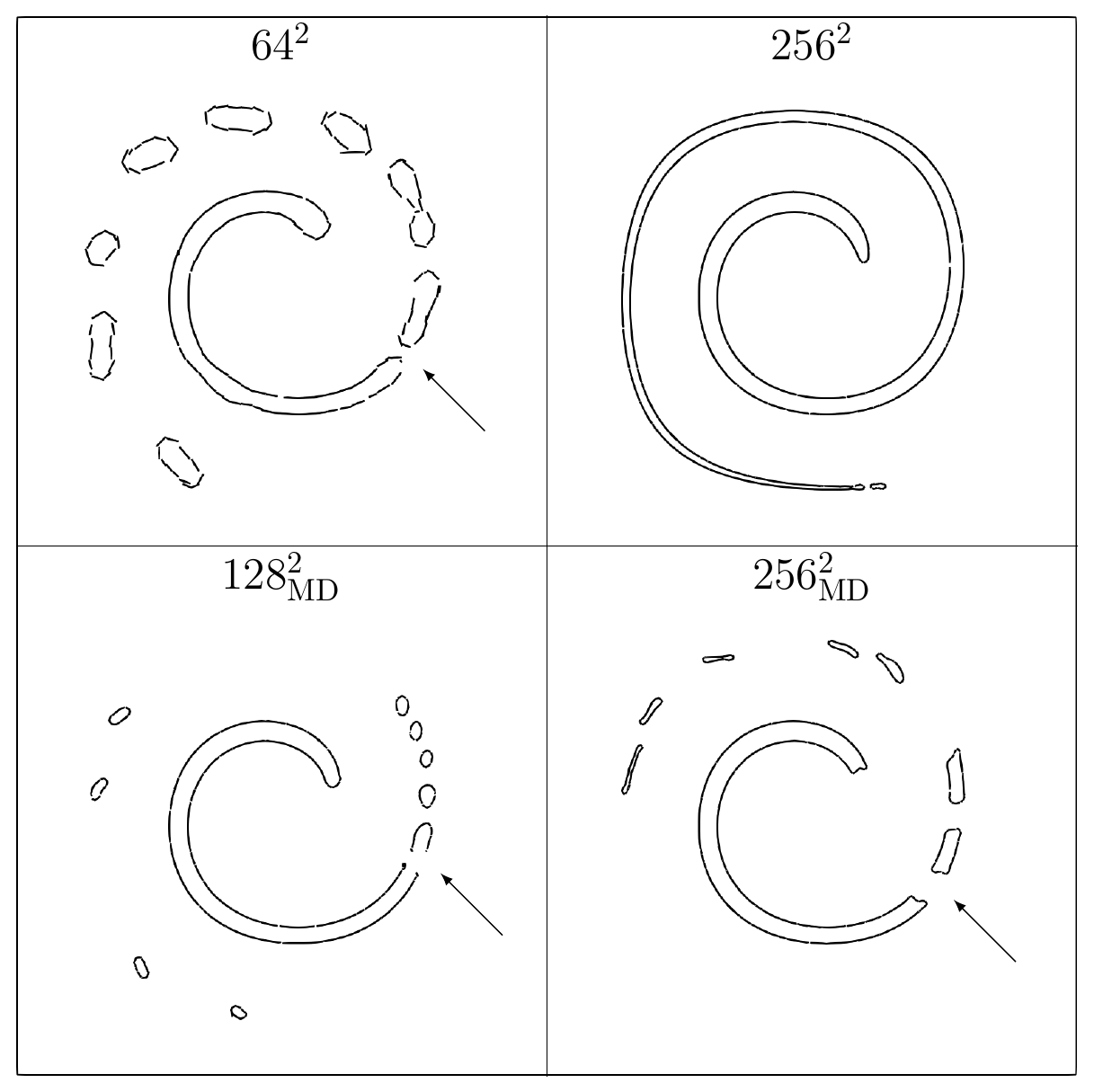}
  \caption{$t=2.35$ s.}  
  \end{subfigure}
  \begin{subfigure}[b]{0.45\textwidth}
  \includegraphics[width=\textwidth]{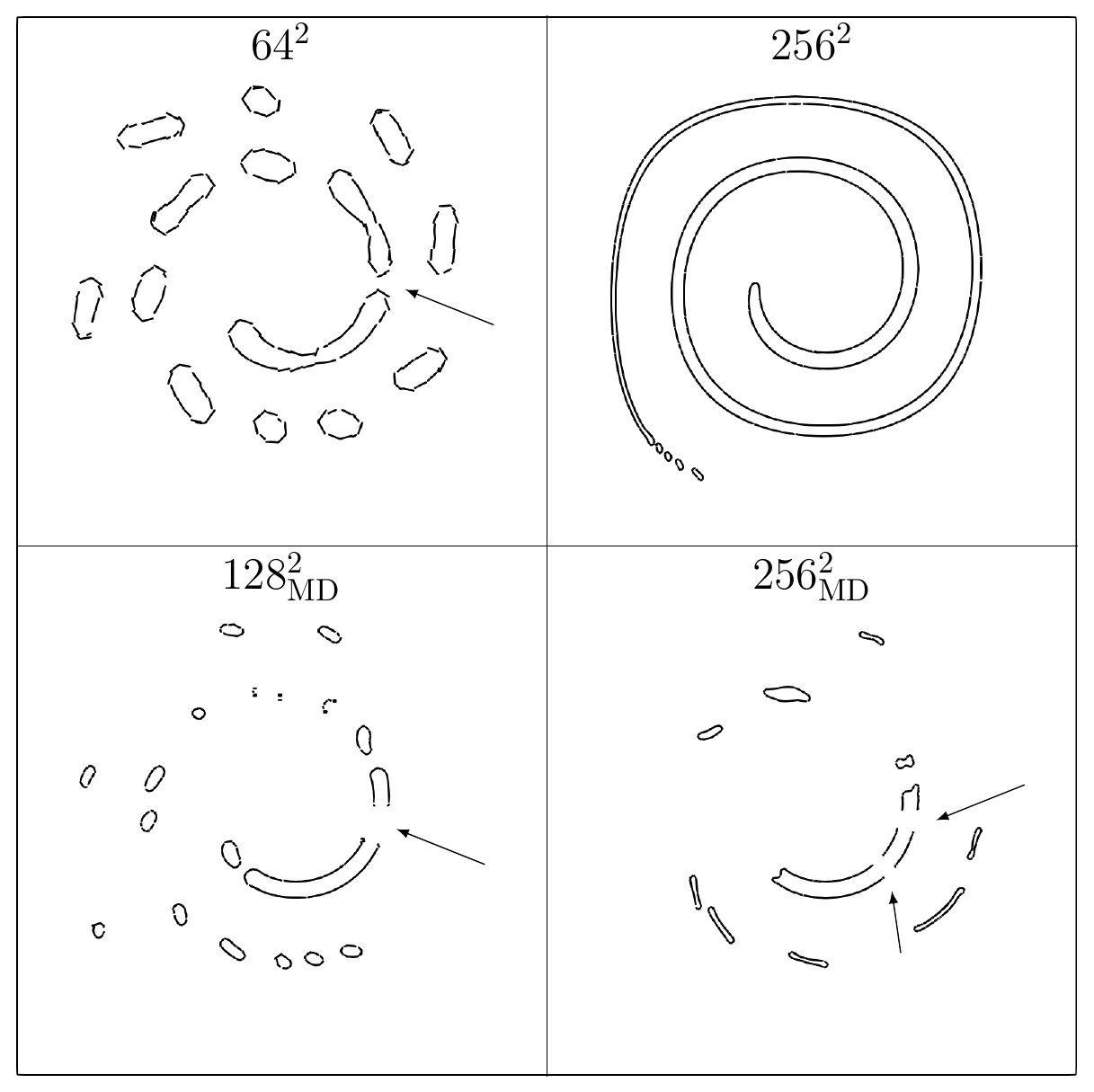}
  \caption{$t=2.95$ s.}  
  \end{subfigure}
  \caption{Position of the interface at various instants for the single vortex test. 
  In each frame the following cases are shown: 
  $64^2$ (top left) and $256^2$ (top right) grids without controlled perforation,
  $128^2$ (bottom left) and $256^2$ (bottom right) grids with manifold death.
  The arrows indicate the position of the latest breakup.
  }
  \label{imm:vortex}
\end{figure}
The single vortex test has been designed to check the ability of interface tracking methods
when the reference phase undergoes large deformations and thin structures of thickness
of the same order of the grid size are involved, see \cite{bell1989second,aulisa2003geometrical}.
For this test, the surface tension is set to zero.
The divergence-free velocity $\mathbf{u} = (u_x , u_y )$
is obtained from the following stream function
\begin{equation}
 \psi = \frac{1}{\pi} sin^2 (\pi x) sin^2 (\pi y) \,,
\end{equation}
as $u_x = \partial \psi/\partial y$ 
and $u_y = -\partial \psi/\partial x$.
The domain is a unit square with the bottom left corner placed in the origin
and a circular droplet with radius $r = 0.15$ is centered at $(0.5,0.75)$.
On the sides of the box, homogeneous Dirichlet boundary conditions are imposed.
The rotation ends at $T=3$ s.

We use this simple test to check the first step of the manifold death algorithm,
namely the detection of thin structures where the artificial breakup occurs.
If we perform the simulation on a coarse mesh
we observe a certain amount of artificial breakup events and, if we move to finer grids,
we expect to obtain fewer breakup events, 
as the filament remains larger than the grid size for a longer time.
Since we aim to tackle the problem of the non-convergence upon grid refinement,
we first show that our method can replicate on finer grids (in a controlled fashion)
what happens numerically on a coarse grid taken as reference.
In other words, we check if the manifold death method is able to identify and perforate
the ligament on finer grids as soon as it reaches a thickness equal
to the reference coarse grid size.
To do so, the critical ligament thickness $h_c$ used in the algorithm
has to be equal to the grid size $\Delta_l$ of the coarse mesh.
This requires 
$\Delta_l=h_c\approx3\Delta_{l+1}$, where $\Delta_{l+1}$ indicates the finer grid size. 
However, using the standard multilevel, $\Delta_l=2\Delta_{l+1}$ holds true and 
since Basilisk uses quadtree grids, which restricts the number of cells per side to powers of two,
the simpler way to satisfy this requirement is to increase the coarse grid size $\Delta_l$
by taking a square domain of size 1.5 
(note that this would be equivalent to using a $47^2$ grid on the unit square).

We show in Figure \ref{imm:vortex} the interface at
different grid resolutions with and without manifold death.
The four sub-pictures correspond to different times. 
In the top row of each picture we report the interface with no manifold death
for the $64^2$ (used as reference) and $256^2$
(only reported as a comparison) grids,
while in the bottom row the manifold death method is used
with constant $h_c=0.023\rm{m}=3\Delta_{128}$ on the $128^2$ and $256^2$ grids.
Note that, for clarity, the smallest droplets have been removed.
The criterion used for the mesh adaptation is the error
on the volume fraction field $C$ with a threshold equal to $1\cdot10^{-5}$.
The frequency of holes creation is $t_h=0.05$ s and the maximum number
of holes per manifold death call is $n_h=10$.

Using the standard VOF the breakup of the thin filament is clearly grid dependent,
since with the coarser ($64^2$) grid at the end of the simulation the reference phase
is made of a set of droplets,
while with the finer ($256^2$) one the thin filament is still well resolved 
and only few droplets have been created near the end of the tail.
The validation of the manifold death method, and in particular the detection of thin structures,
can be done qualitatively by comparing the position of the most recent breakup event
(indicated by the arrows) for the two cases with manifold death and for the $64^2$ case
without controlled perforation.
By controlling the breakup, we have replicated on finer grids
what happens numerically on a coarser grid,
showing that the method accurately detects and perforates thin structures
of given thickness independently of the grid resolution used.
\subsection{Secondary atomization}
Secondary atomization refers to the physical situation where the droplets
originated from the primary atomization of a jet, due to the interaction 
with the ambient high speed gas flow, deform and fragment.
Here, we study the evolution of a single droplet that experiences an impulsive
acceleration of the surrounding gas.
The droplet is initially at rest, surrounded by fluid with zero velocity.
At $t=0$ an impulsive velocity $U_0$ condition is imposed on the left boundary.
The drop then stretches and deforms into a film whose shape resembles that of a bag.
As the bag inflates, its thickness decreases until holes appear and the bag breaks up
bursting the bag into a spray.
An extensive analysis of the different fragmentation regimes that
the droplets may experience has been carried out in 
\cite{pilch1987use,guildenbecher2009secondary,marcotte2019density}.
The goal of this paper is not to further investigate the secondary atomization problem,
instead we want to show the improvements in terms of grid convergence 
that can be attained using the manifold death method.
To this aim, we select a single case from \cite{marcotte2019density}
with the intent to show that we are able to control the sheet breakup,
while in the standard VOF simulations it happens numerically
as soon as the sheet thickness reaches the minimal grid size.

The computational domain is a two-dimensional axisymmetric channel, represented by a square
of side $15 R_0$, where $R_0$ is the droplet initial radius used in the adimensionalization.
The dynamics of the droplet can be uniquely determined by the
Reynolds $\Re=\rho_G U_0 R_0/\mu_G$ and Weber $\We=\rho_G R_0 U_0^2/\sigma$ numbers,
the ratio of densities $r_d=\rho_L/\rho_G$ and the ratio of viscosities $r_v=\mu_L/\mu_G$.
For this test we take 
$\Re=1090$, $\We=7.5$, $r_d=1110$ and $r_v =90.9$.
We perform the simulations on three fine grids
$8192^2$, $16384^2$ and $32768^2$
(corresponding to the levels of refinement 13, 14 and 15) 
with up to 2000 cells per droplet radius.
Adaptation is performed using a threshold
of $10^{-5}$ and $5\cdot10^{-5}$ on the velocity $\mathbf{u}$ and volume fraction $C$ errors, respectively.
For this test case we set $t_h=0.1$ and impose that only one hole
is done during the whole simulation.   
To reduce the effects of the long initial transient
where the droplet stretches and forms the forward-facing bag,
we start the simulations on the finer grid by using a snapshot 
obtained at lower resolution.
\begin{figure}[htb!]
 \centering
  \includegraphics[width=0.98\textwidth]{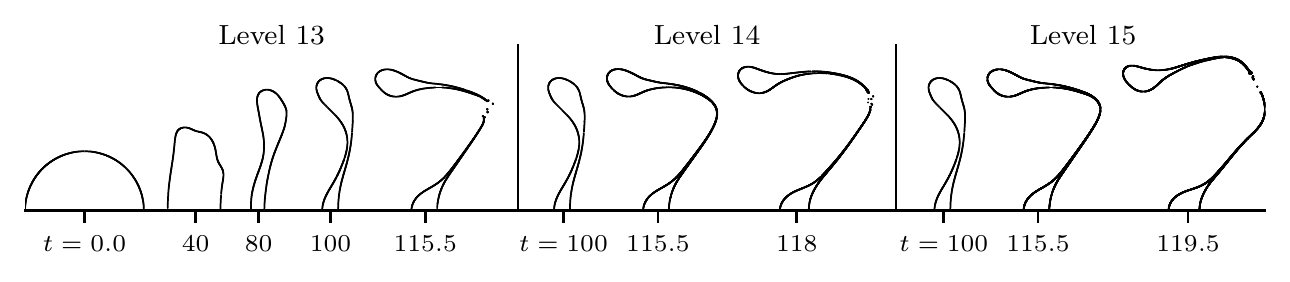} 
\caption{Drop deformation and breakup without manifold death (non controlled)
for three grid resolutions.
}
 \label{imm:hist}
\end{figure}
Figure \ref{imm:hist} shows the evolution of the droplet until its breakup for the different grids.
The droplet deforms and a forward facing bag is formed. 
The breakup occurs at $t=115.0$ at the lower resolution 
and at $t=119.2$ when using the finer grid, see Table \ref{tab:breakup}.
The evolution is similar to the one in \cite{marcotte2019density},
apart from the bulge near the symmetry axis that
was not observed in the previous work.

\begin{figure}[htb!]
 \centering
    \includegraphics[width=0.49\textwidth]{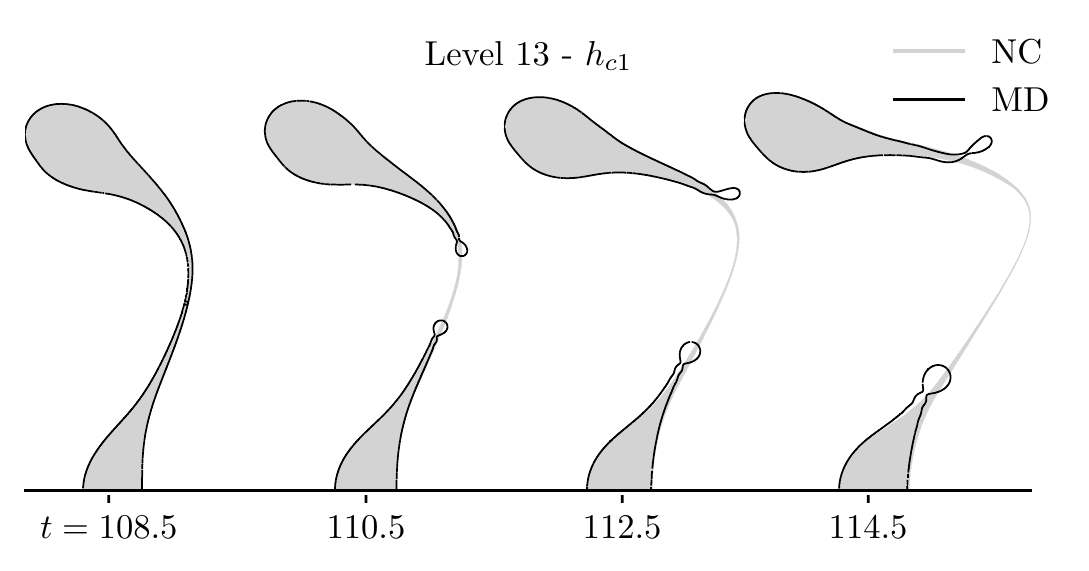}
    \includegraphics[width=0.49\textwidth]{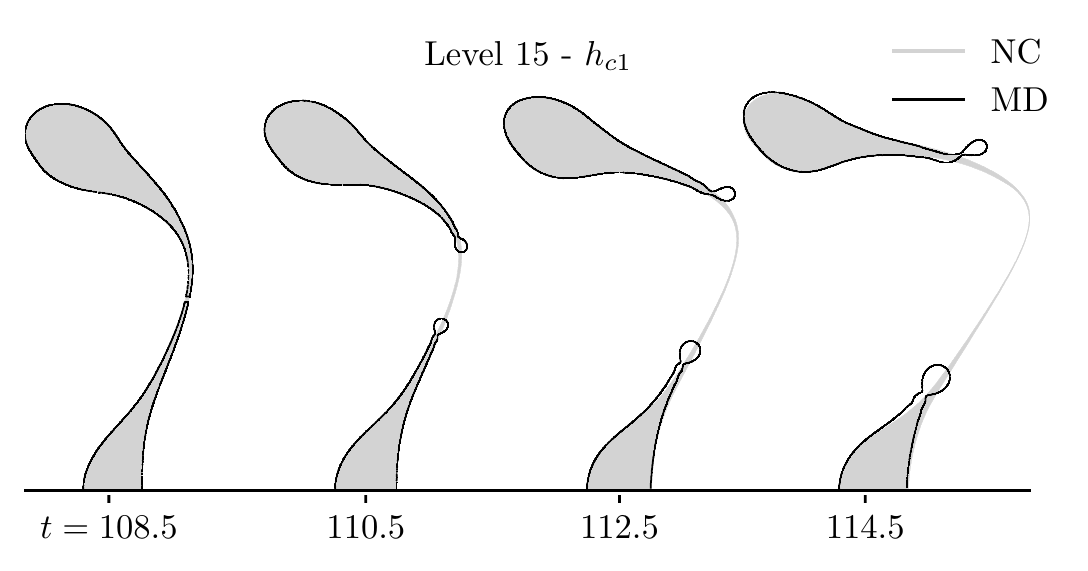}
    \includegraphics[width=0.49\textwidth]{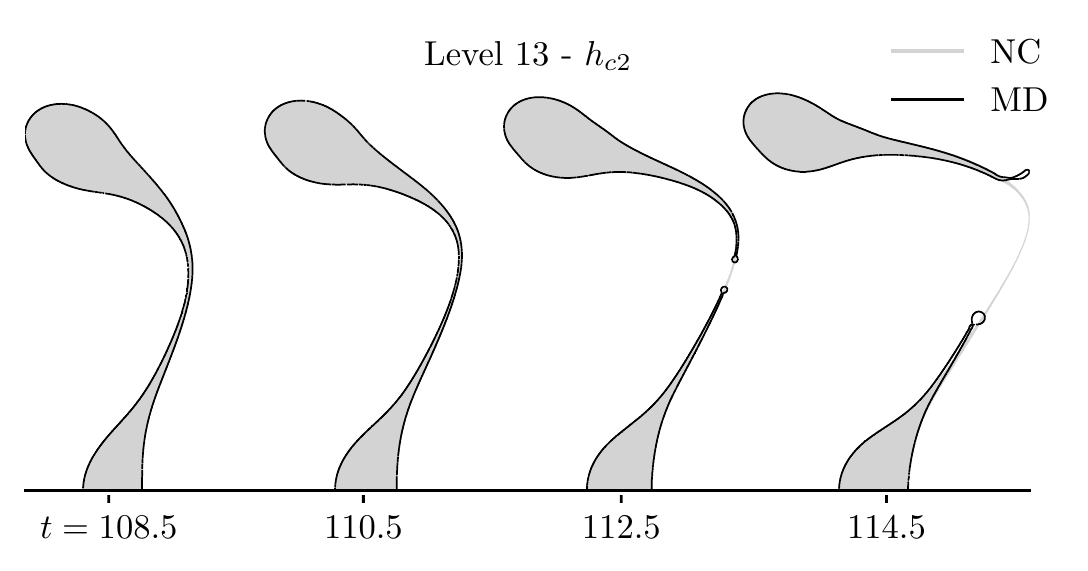}
    \includegraphics[width=0.49\textwidth]{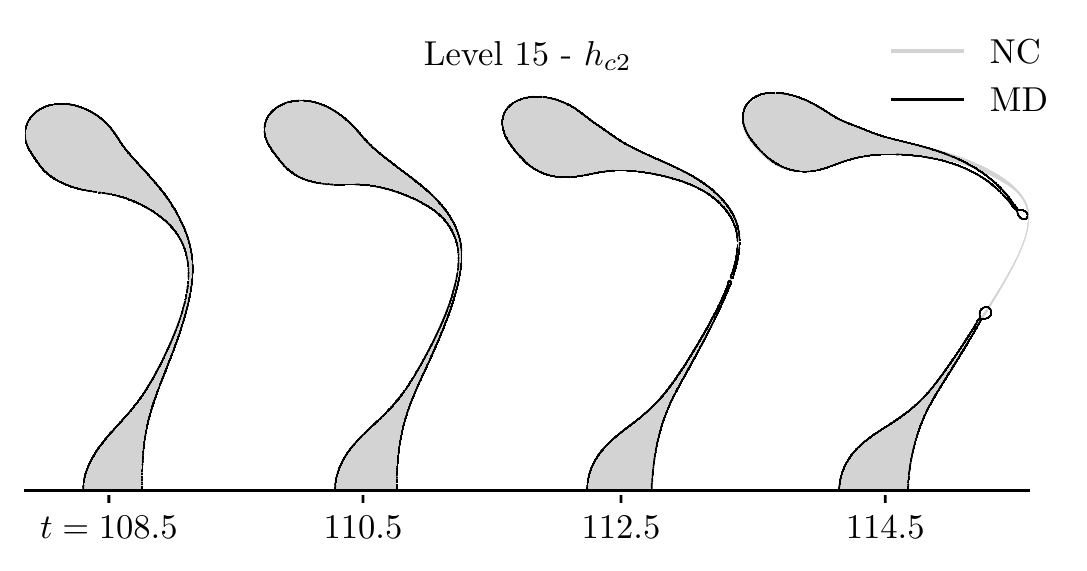}    
    \includegraphics[width=0.49\textwidth]{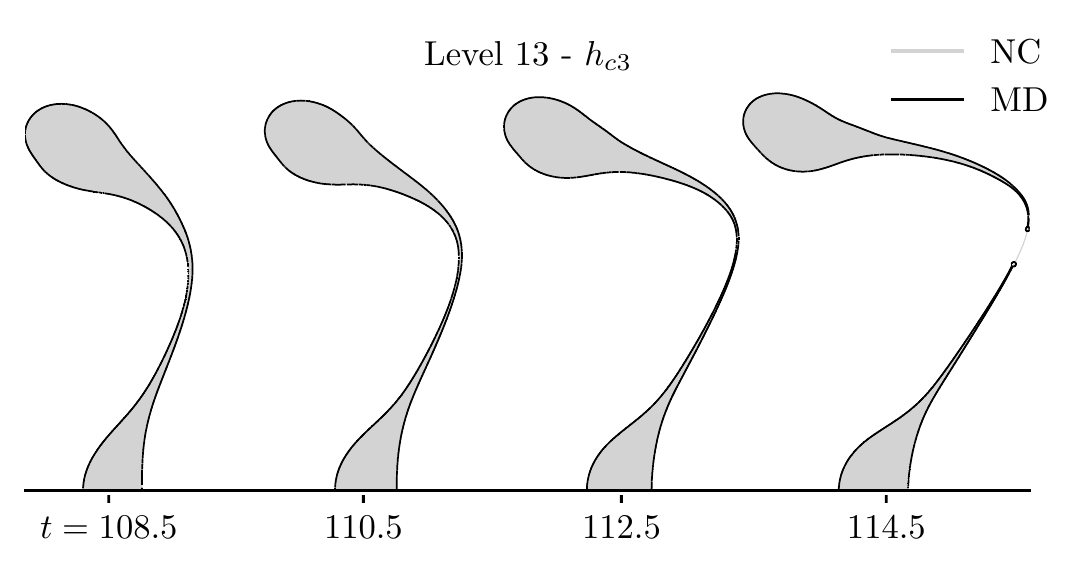} 
    \includegraphics[width=0.49\textwidth]{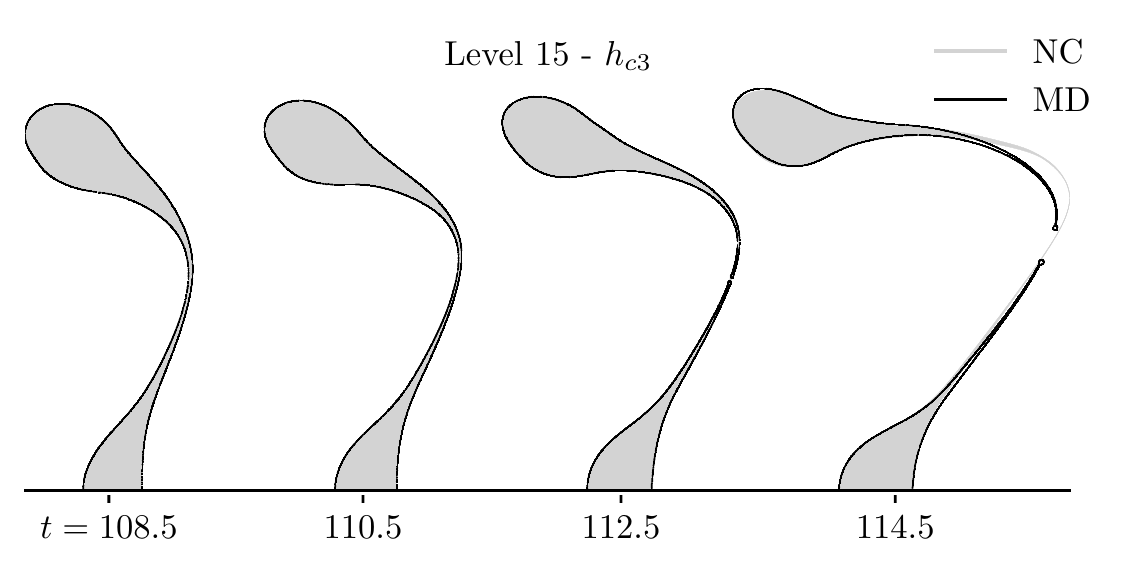} 
  \caption{Drop deformation using the $8192^2$ (Level 13, left) and $32768^2$ grids (Level 15, right).
  The black line shows the position of the interface when the manifold death (MD) method is used,
  while the gray background shows the droplet position with non controlled (NC) breakup.
  The critical thickness ranges from $h_{c1}=3\Delta_{2048}$ (top) to
  $h_{c2}=3\Delta_{4096}$ (center) and $h_{c3}=3\Delta_{8192}$ (bottom).
  $\Delta_{N}$ is equal to $L/N$, where $L$ is the domain size and $N$
  the number of cells per side.
 }
  \label{imm:manifold}
\end{figure}

The breakup of the droplet is shown in Figure \ref{imm:manifold} 
for the coarsest and finest grids.
The black line indicates the interface when the manifold death (MD) method is used, while the gray background shows
the position of the deformed droplet in the case with non controlled breakup (NC) at the same mesh resolution.
From top to bottom, we report the results obtained 
reducing the critical thickness $h_c$ 
at which thin sheets are detected and perforated. 
In the top row we set $h_{c1}=3\Delta_{2048}=0.022R_0$,
in the middle one $h_{c2}=3\Delta_{4096}$ and, finally,
in the bottom one $h_{c3}=3\Delta_{8192}$.
Table \ref{tab:breakup} lists the time at which the first breakup event happens.
By reducing the value of $h_c$ the bag breaks later 
no matter the mesh resolution used for the solution of the Navier-Stokes equations.
Also, with a given critical thickness and comparing the results for the two resolutions shown, 
the breakup occurs in the same spot and almost at the same time, 
indicating that we successfully managed to control the breakup
making it almost grid independent.
To summarize, in all the cases with manifold death, 
the breakup happens earlier
than in the reference case without controlled perforation
and the anticipation time depends on the critical thickness $h_c$.

\begin{table}[htb!]
\centering
   \begin{tabular}{cccc} 
     \hline\hline
     Test case & $8192^2$ grid & $16384^2$ grid & $32768^2$ grid   \\
     \hline
     Non controlled  & 115.0 & 117.5& 119.2 \\
     $h_{c1}=3\Delta_{2048}$  & 108.5 & 108.5 & 108.5 \\
     $h_{c2}=3\Delta_{4096}$  & 112.2 & 112.4 & 112.5\\
     $h_{c3}=3\Delta_{8192}$  & 114.0 & 114.1 & 114.2\\
     \hline\hline
  \end{tabular}
  \caption{Adimensional time at which the first breakup event occurs in the non controlled case and with three different critical thickness $h_c$.}
  \label{tab:breakup}
\end{table}
\subsection{Phase inversion problem}\label{sec:phaseInv}
\begin{figure}[htb!]
 \centering
  \includegraphics[width=0.4\textwidth]{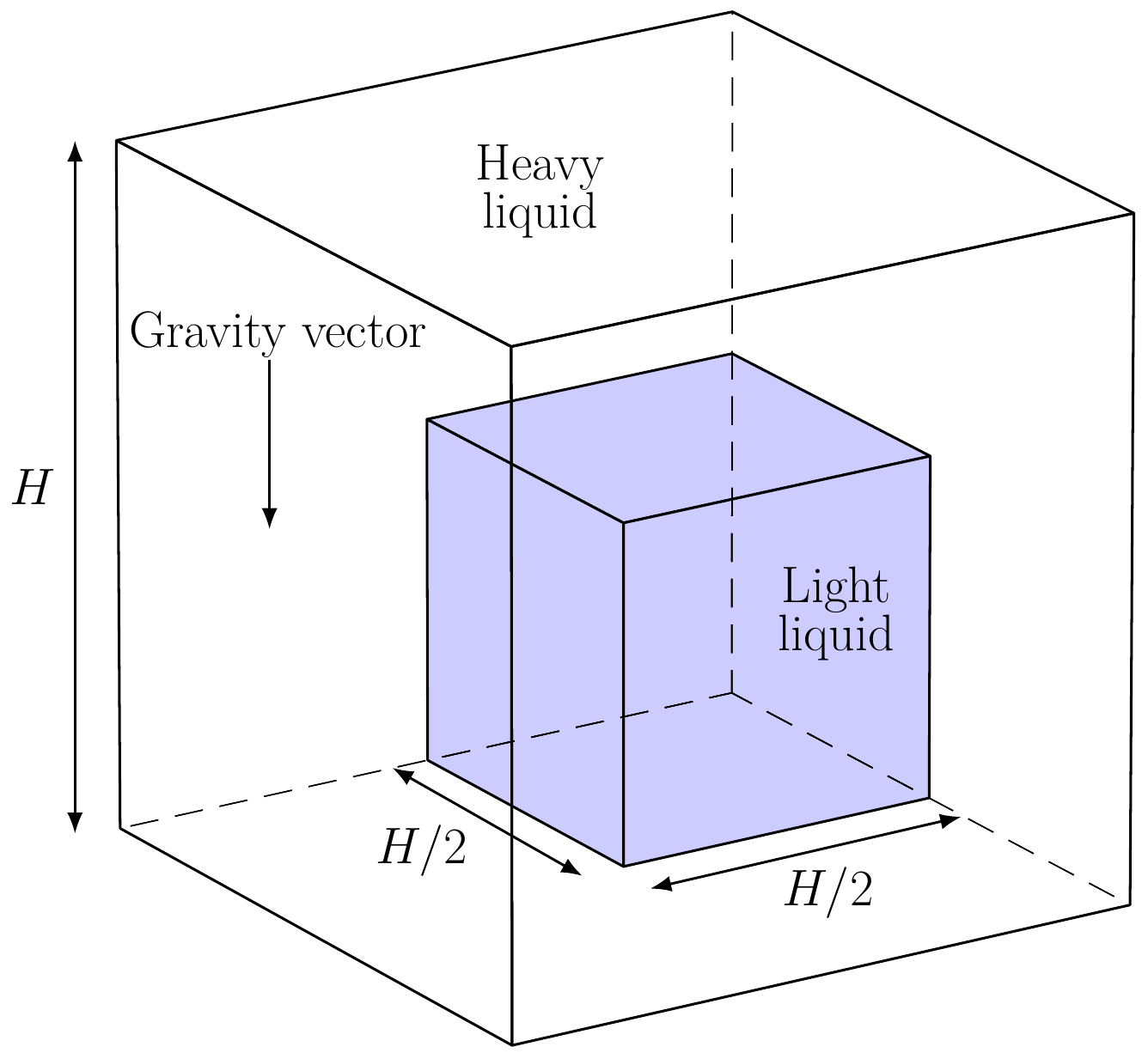}
  \caption{Configuration for the phase inversion test.} 
  \label{imm:domPh}
\end{figure}
The situation of the phase inversion consists of two fluids, initially at rest,
with the lighter one placed in the bottom of a box surrounded by the heavier one,
see Figure \ref{imm:domPh}. 
The outer cubic box has size $H$, while the inner one $H/2$.
On the outer walls we impose a free-slip boundary condition,
so that the normal velocity is zero and the tangential components obey
a symmetry condition. 
A $\pi/2$ static contact angle is imposed on the walls and
the gravitational acceleration is $\mathbf{g} = (0, -9.81, 0)$. 
Due to gravity the lighter fluid raises and in the end
will occupy the top part of the box. 

In between the simple initial and final configuration, it is possible
to observe many breakup and coalescence events, which in standard VOF simulations
are due to both numerical and physical aspects.
In particular, the role that sheets breakup plays in the lack of convergence
of enstrophy has been recently pointed out in \cite{sayadiconvergence}.
The satisfying grid convergence of primary moments, such as kinetic and mechanical energy, 
has already been shown in literature \cite{Tcase15,saeedipour2021toward,estivalezes2022phase}
and therefore here we do not provide a discussion about this aspect.
In this work, however, we show that the manifold death method prevents
the numerically induced breakup of thin structures
and, as a consequence, the grid convergence of enstrophy
and droplets size that plagues many of the literature cases is improved.

In this study, we consider moderate values of the Archimedes and Bond numbers,
aiming to perform a true DNS multiphase simulation, with an accurate resolution of the fluid flow.
The geometrical dimensions and the physical properties of the two fluids
are reported in Table \ref{tab:physProp}. 
Note that the properties of the lighter fluid resembles those of oil
and those of the heavier one of water. 
\begin{table}[htb!]
\centering
   \begin{tabular}{ccccccccc} 
     \hline\hline
     $L$ & $\mu_w=\mu_o$ &$\rho_w$&$\rho_o$&$\sigma$&$g$&  Ar$^{1/2}$& Bo &  Ca \\
     \hline 
     (m)&(Pa s)&(kg m$^{-3})$&(kg m$^{-3})$&(kg s$^{-2})$&(m s$^{-2}$)&-&-&-\\
     \hline 
     0.1   & 0.01958 & 1000 &  900& 0.01533&9.81& 1600 &  640 &  0.4 \\ 
     \hline\hline
  \end{tabular}
  \caption{Physical properties and dimensionless numbers.}
  \label{tab:physProp}
\end{table}

We introduce the following definitions for adimensional numbers and quantities
\begin{align}
\begin{aligned}
  \Ar = \frac{\rho_w(\rho_w-\rho_o)gL^3}{\mu^2}\,, \qquad
  \Bo = \frac{(\rho_w-\rho_o)gL^2}{\sigma}\,,  \\
  \Ca = \Bo/\Ar^{1/2}\,, \qquad t_c= L/U_g=0.4515 \textrm{ s} \,.
\end{aligned}  
  \end{align}
The enstrophy in both fluids is obtained with
\begin{equation}
 E_n=\frac{1}{2}\int_{\Omega} C_n \omega^2 dV \,,
\end{equation}
where $\omega=\nabla \times \mathbf {u}$ denotes the vorticity.
The enstrophy is integrated over the smaller subdomain 
$[0,0.95H]\times[0,0.95H]\times[0,H]$ to neglect
regions where the dynamics is strongly affected by
the presence of the walls of the box, see \cite{sayadiconvergence}.
\begin{figure}[htb!]
 \centering
  \includegraphics[width=0.49\textwidth]{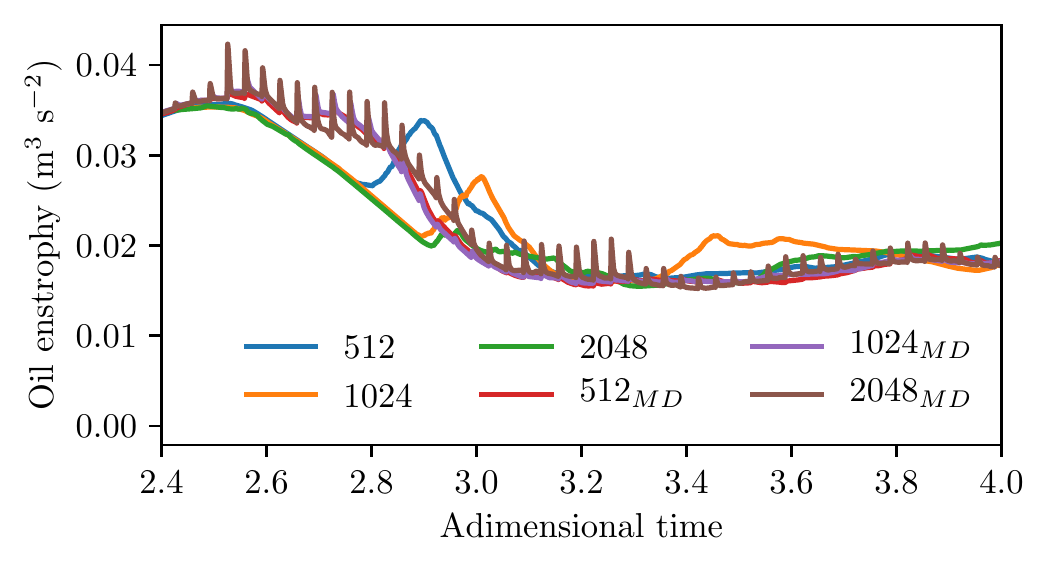}
  \includegraphics[width=0.49\textwidth]{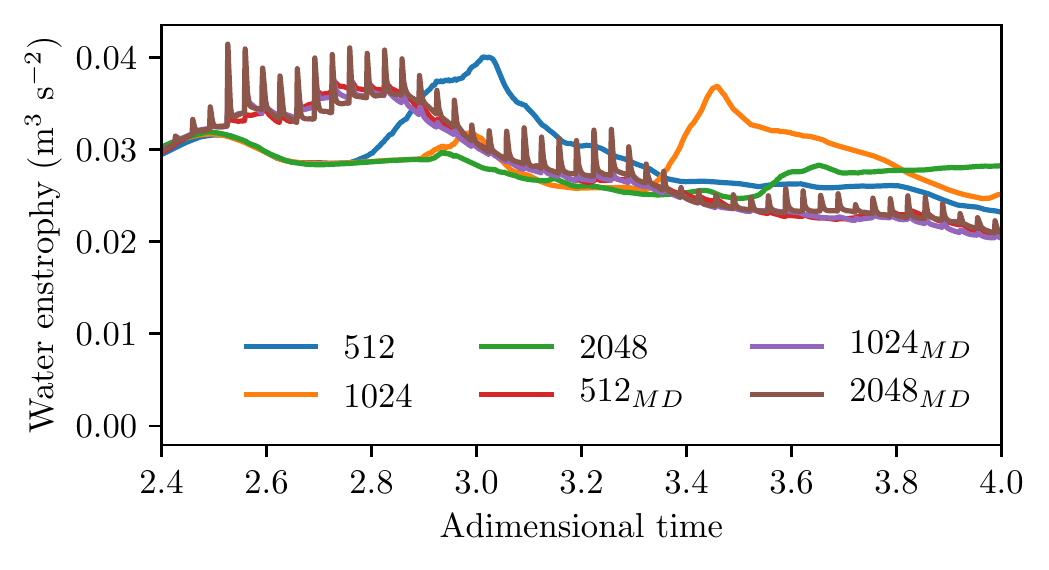}
  \includegraphics[width=0.49\textwidth]{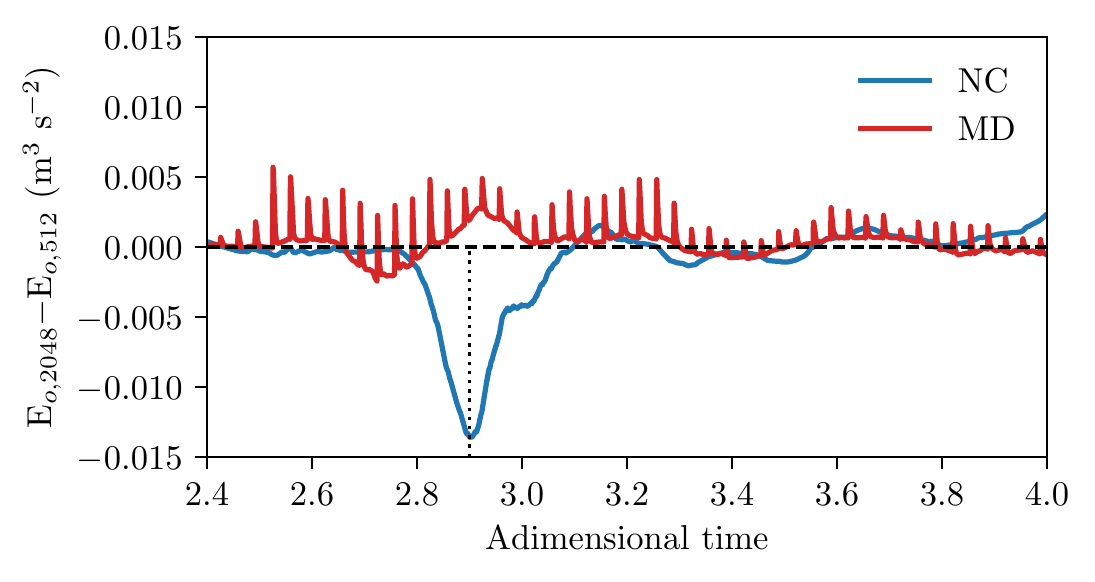}
  \includegraphics[width=0.49\textwidth]{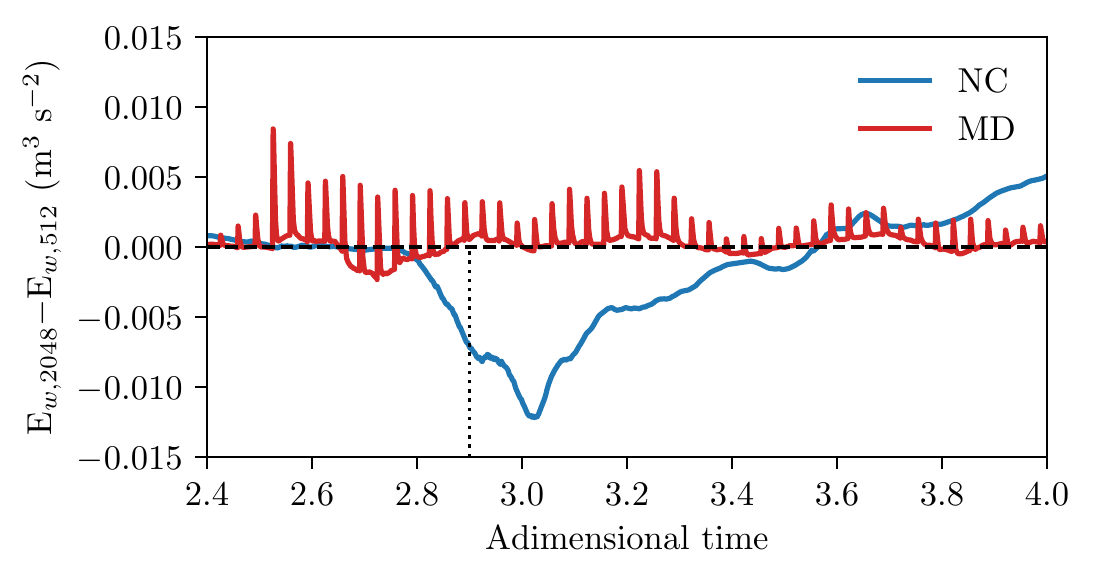}
  \caption{Top row: Oil (left) and water (right) enstrophy.
  Bottom row: Oil (left) and water (right) enstrophy difference between the finest and coarsest grid. 
  The vertical dotted line at $t^*=2.9$ indicates when the thin sheets break in the case without controlled breakup at the coarser resolution.}
  \label{imm:enstrophy}
\end{figure}
\begin{figure}[htb!]
 \centering
  \includegraphics[width=0.4\textwidth]{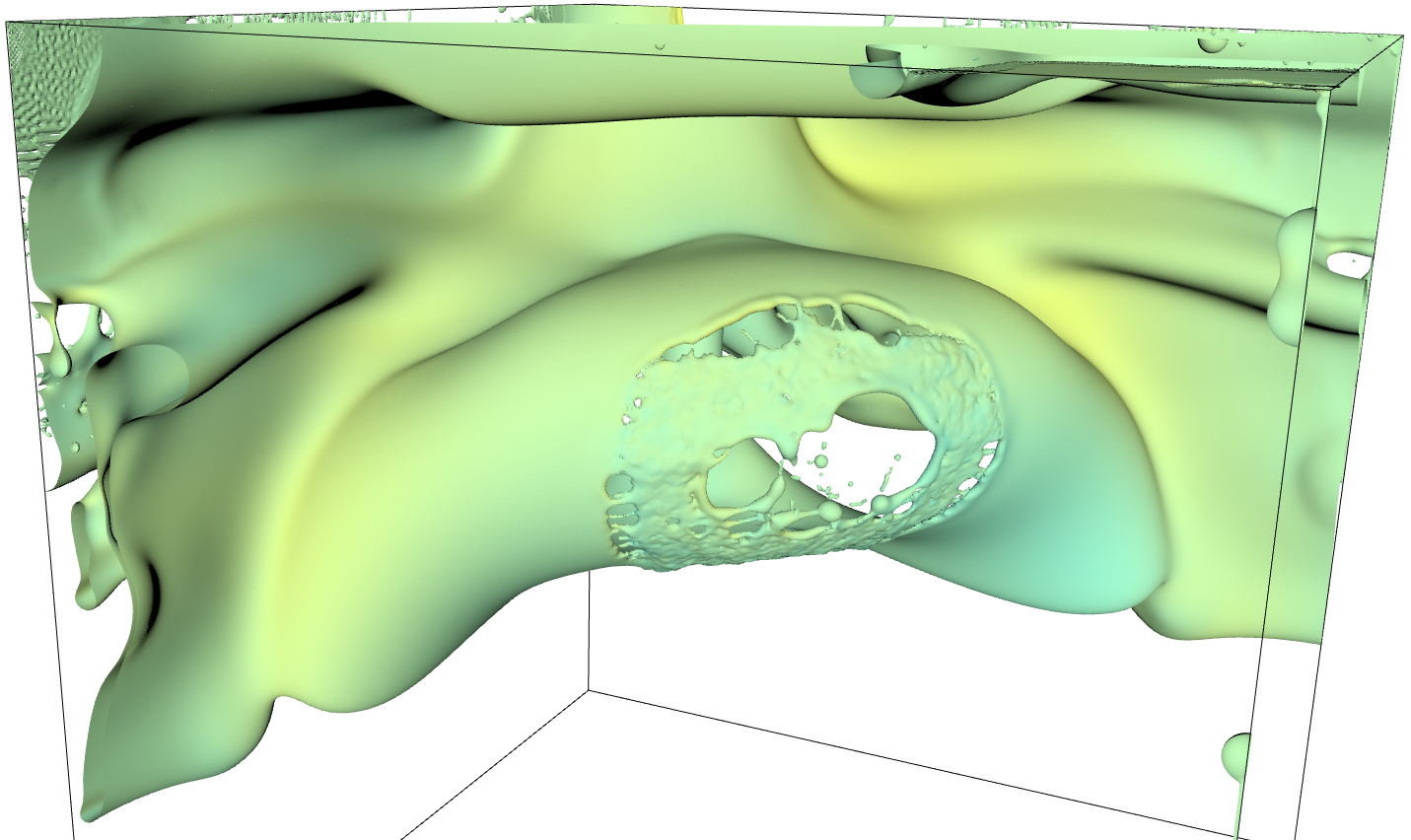}
  \includegraphics[width=0.4\textwidth]{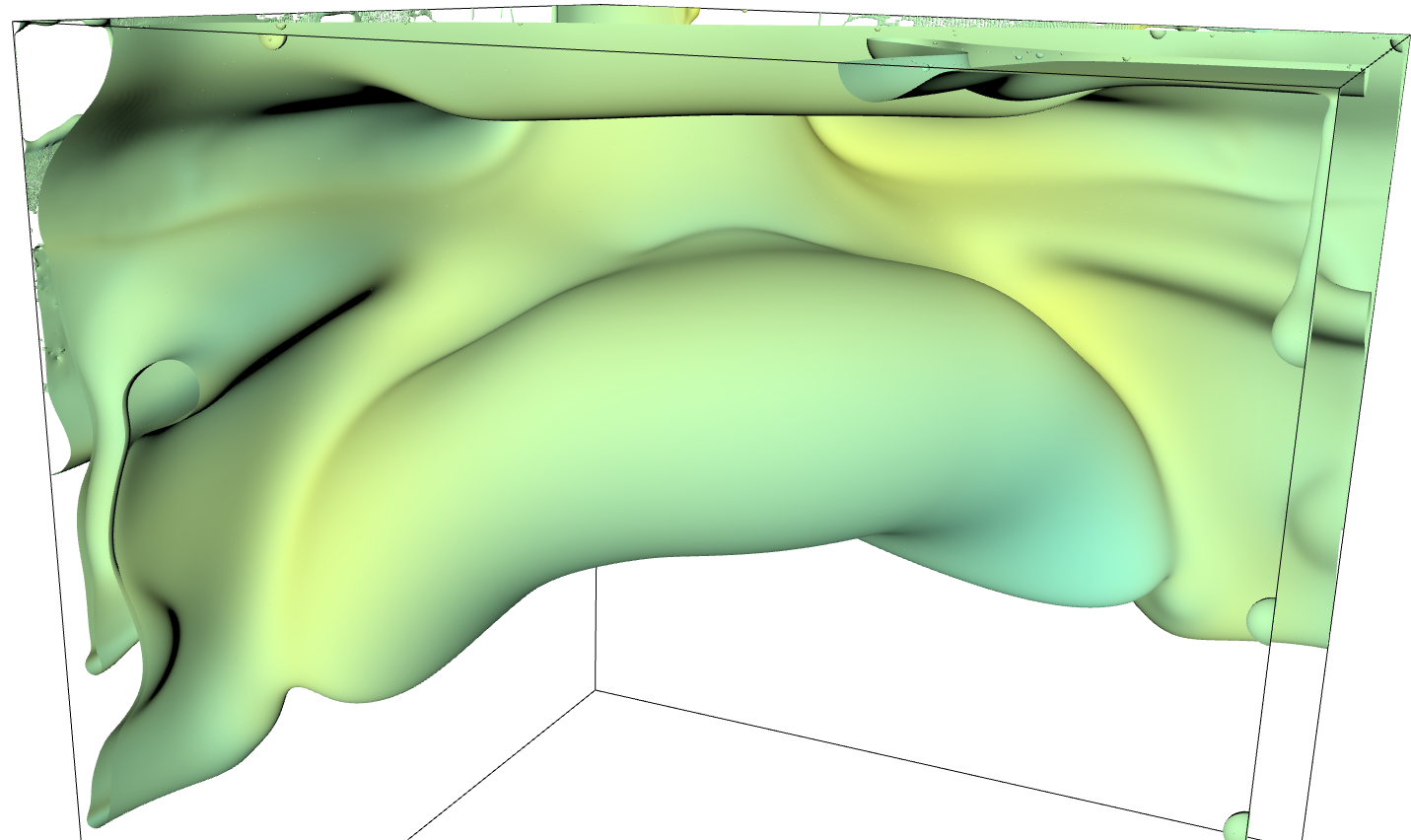}
  \includegraphics[width=0.4\textwidth]{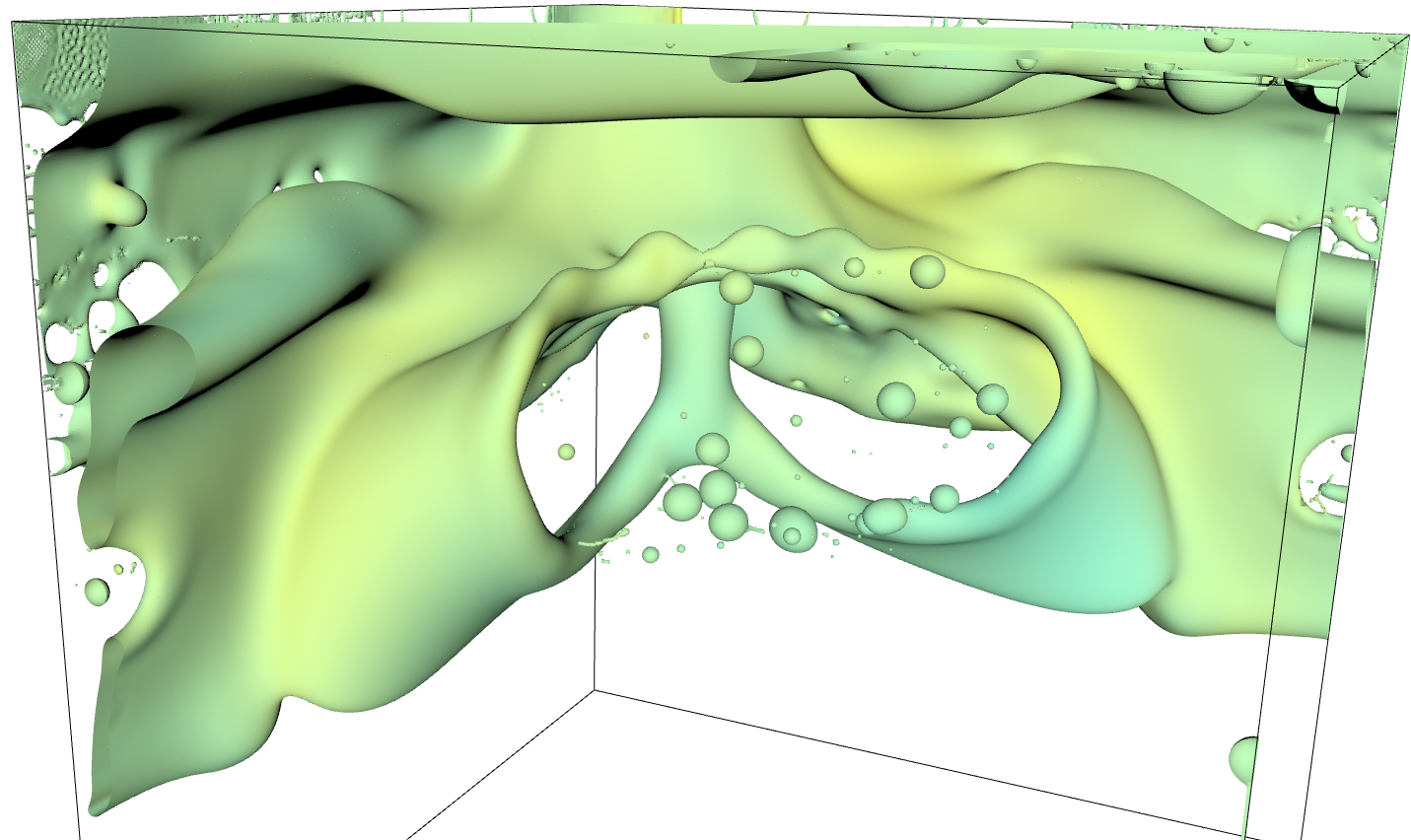}
  \includegraphics[width=0.4\textwidth]{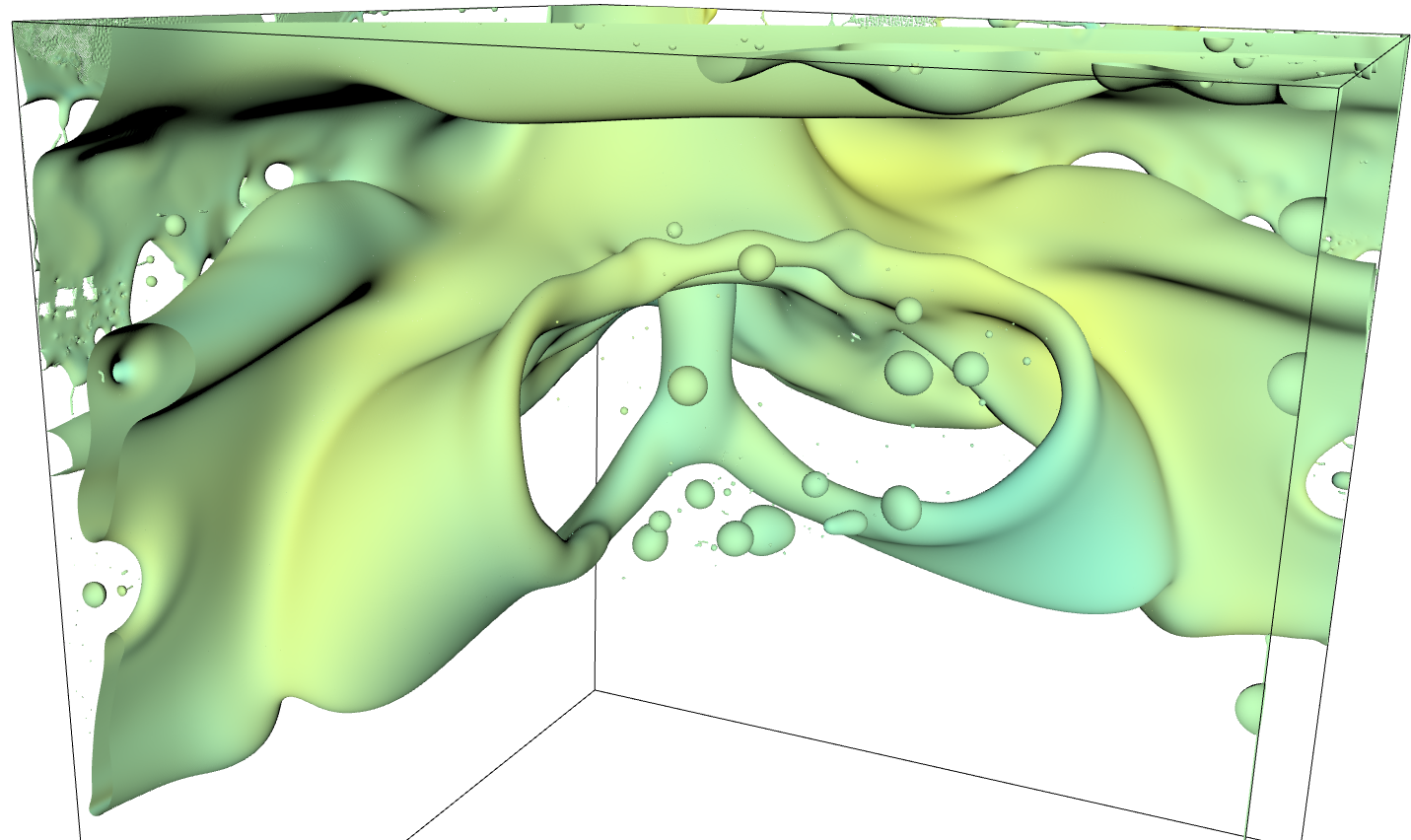}
  \caption{Thin sheet perforation and hole formation in the lighter fluid at $t^*=2.9$. 
  Top: $512^3$ (left) and $1024^3$ (right) grids without controlled perforations. 
  Bottom: $512^3$ (left) and $1024^3$ (right) with manifold death.}
  \label{imm:view}
\end{figure}
\begin{figure}[htb!]
 \centering
  \includegraphics[width=0.245\textwidth]{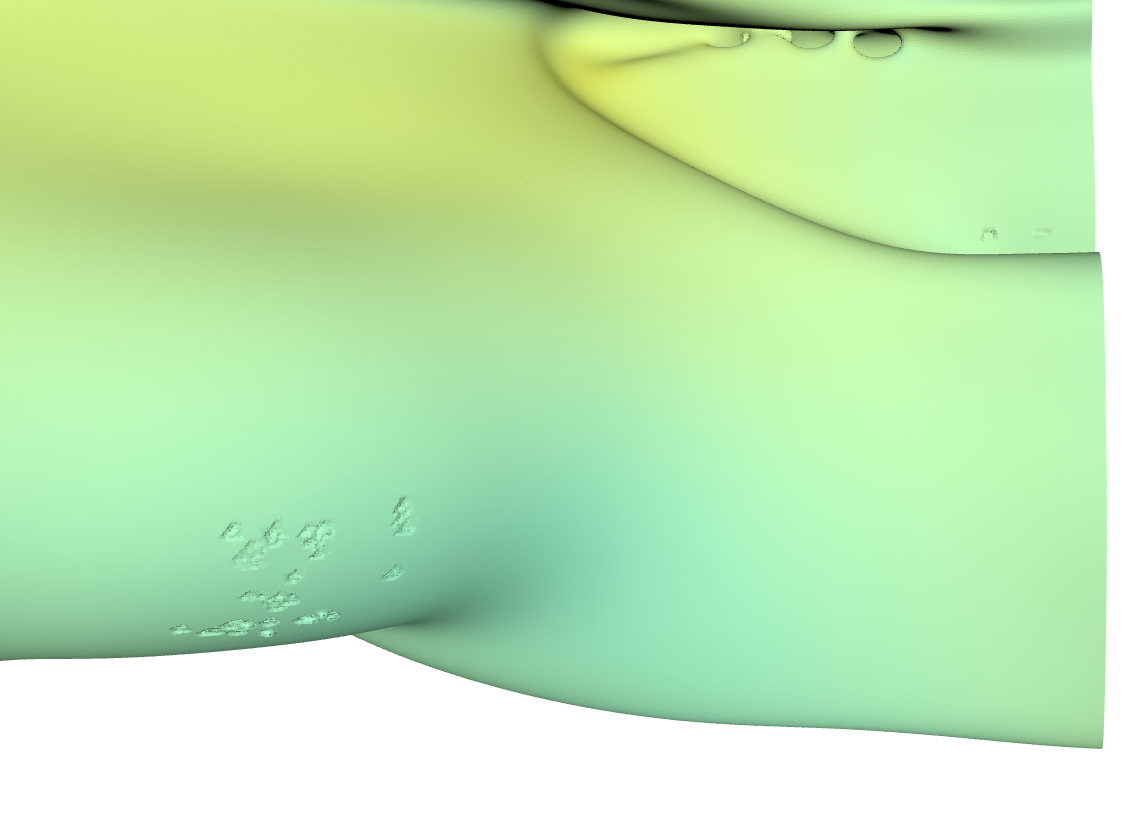}
  \includegraphics[width=0.245\textwidth]{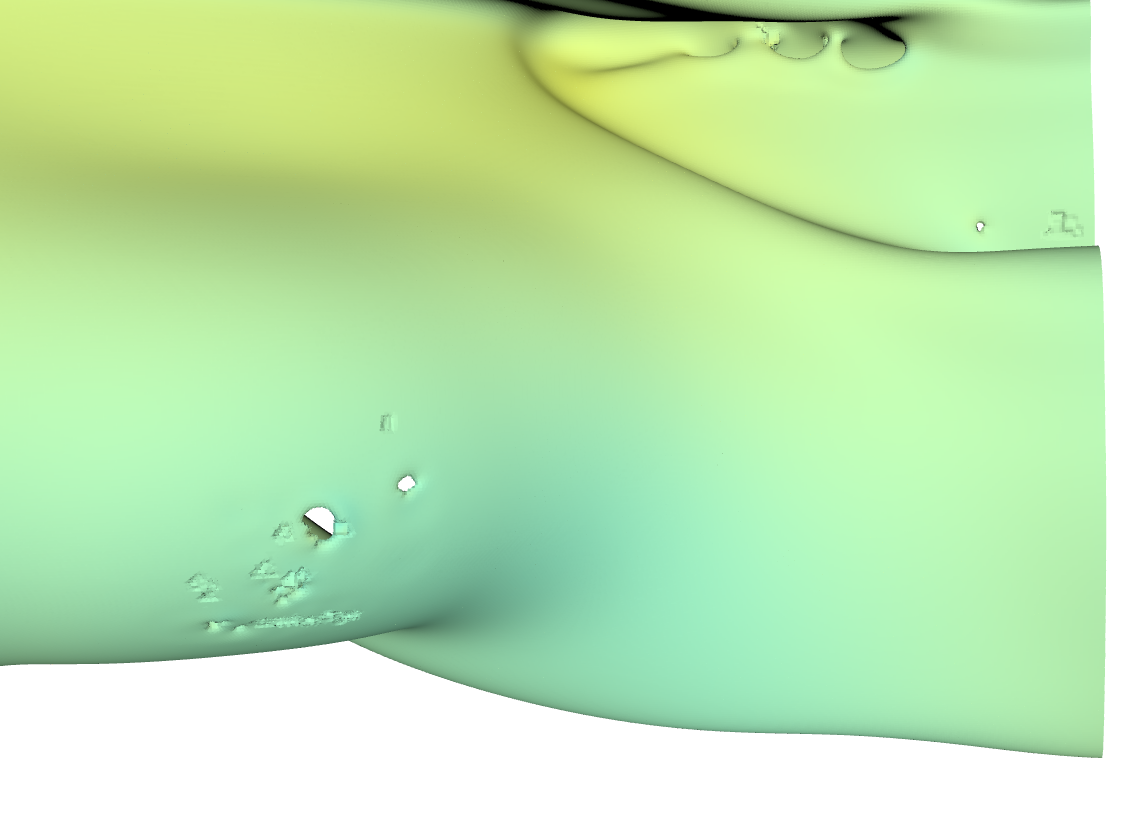}
  \includegraphics[width=0.245\textwidth]{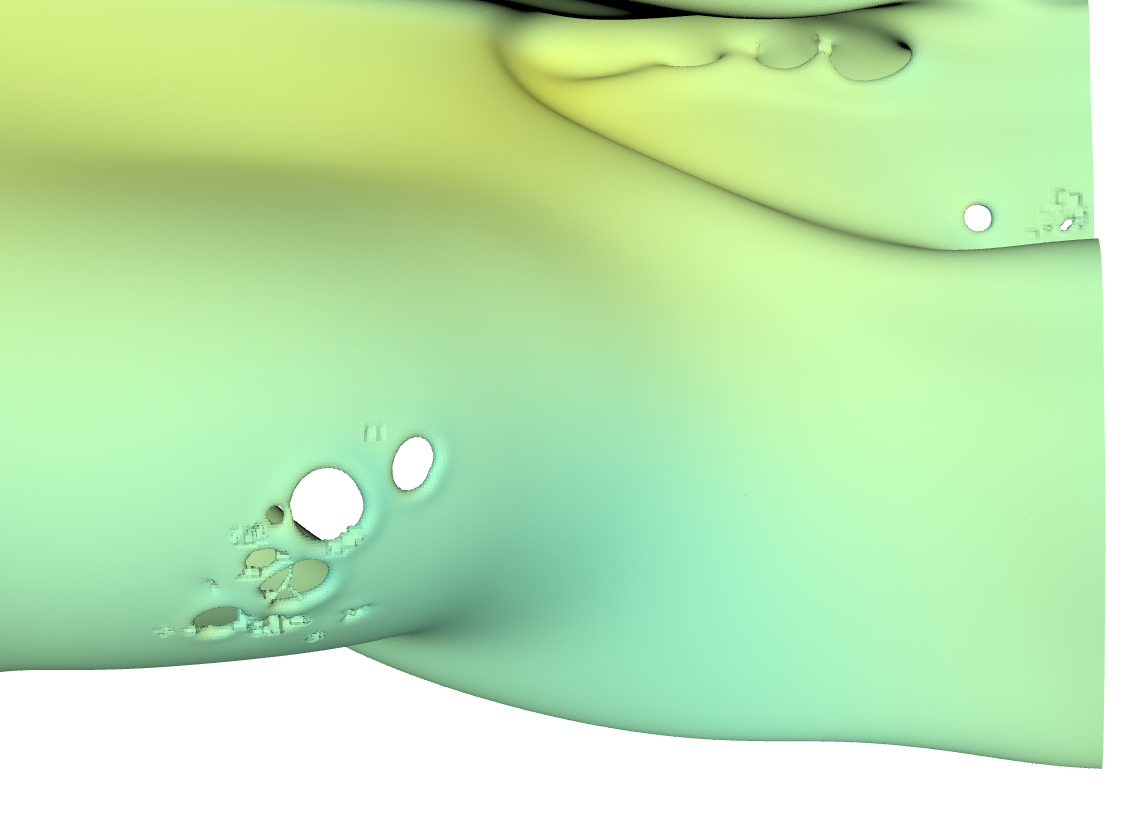}
  \includegraphics[width=0.245\textwidth]{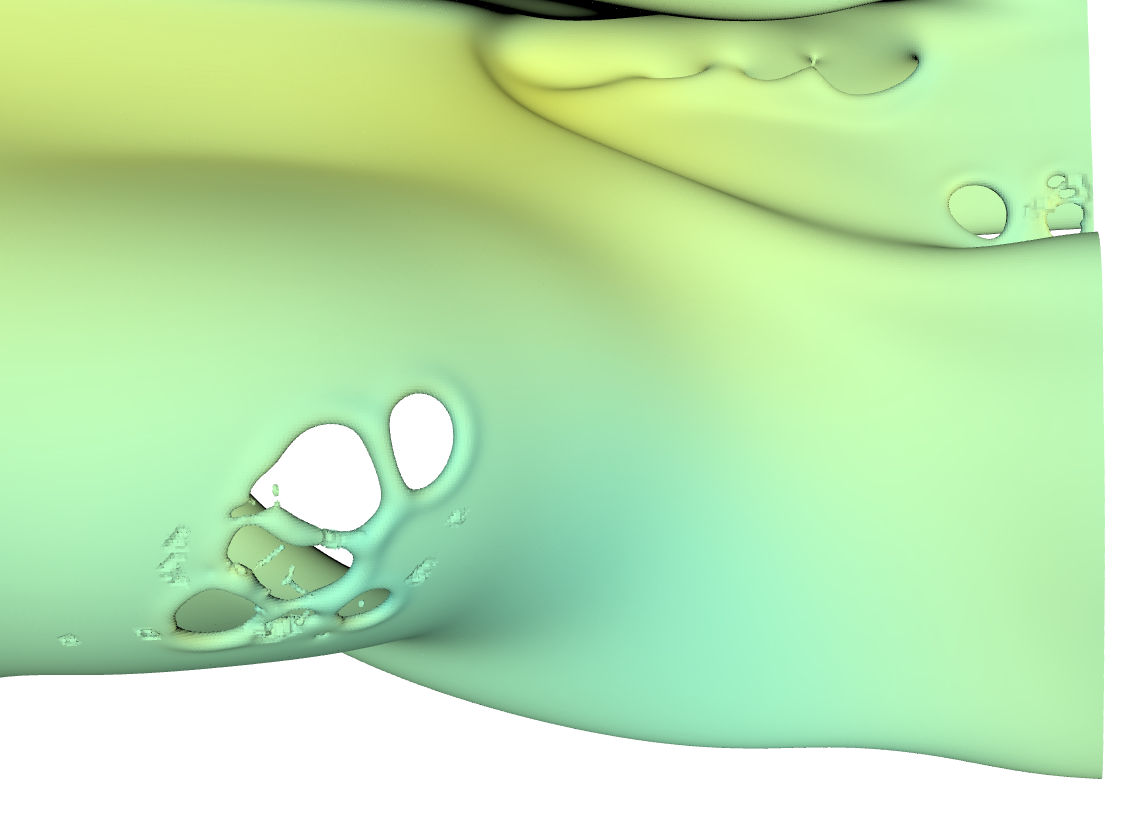}
  \caption{Evolution of the breakup of a thin sheet in the lighter fluid: 
  close-up on the holes expansion. The holes are created by the manifold death algorithm in the first frame.}
  \label{imm:zoom}
\end{figure}
\begin{figure}[htb!]
 \centering
  \includegraphics[width=0.5\textwidth]{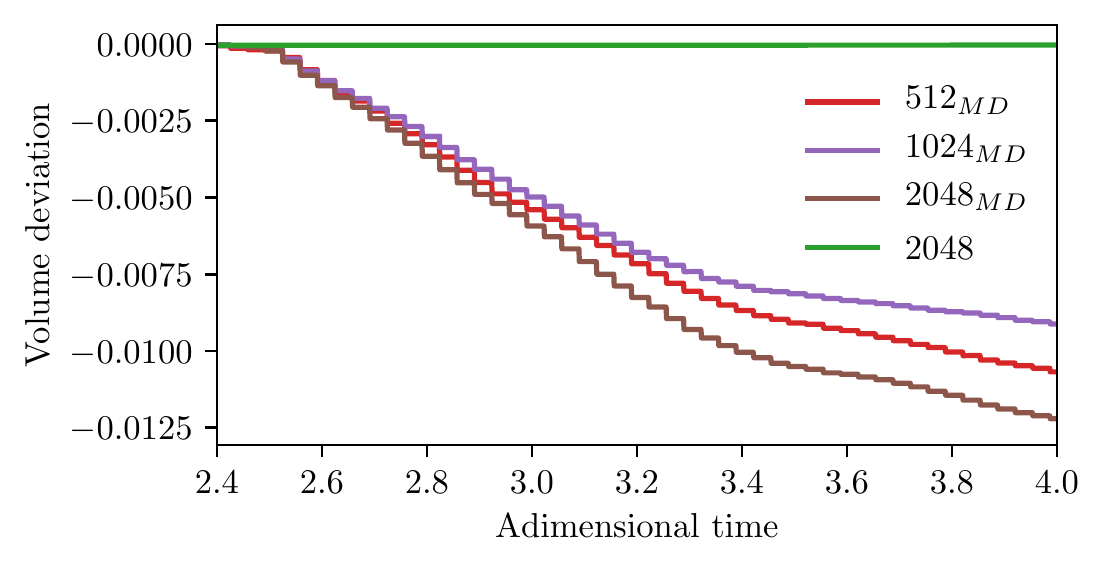}
  \caption{The volume deviation ($V/V_0-1$) of phase 1 (lighter fluid)
  for the three cases with manifold death (MD) 
  and without controlled perforation at the highest resolution.} 
  \label{imm:mass}
\end{figure}

The simulations are carried out on $512^3$, $1024^3$ and $2048^3$ grids
and the constant critical thickness is $h_c=3\Delta_{512}$.
The thresholds used for the refinement criteria are 
$5\cdot10^{-3}$ and $5\cdot10^{-4}$ for the velocity 
and volume fraction errors, respectively.
Concerning the manifold death parameters,
the adimensional time interval between consecutive holes formation is $t^*_h=0.03$
and the maximum number of holes per iteration is $n_h=240$.
In Figure \ref{imm:enstrophy} the improvements on the convergence of the oil enstrophy 
using the manifold death algorithm can be appreciated. 
In the standard case, the enstrophy peaks later
using the finer grids and the local maxima decrease.
Instead, when the manifold death algorithm is used to control the topology changes,
the grid convergence of enstrophy is obtained.
Enstrophy jumps appear when using the manifold death algorithm
and their frequency is clearly $1/t^*_h$.
They are induced by the creation of holes and we believe 
that their amplitude can be significantly reduced 
by creating holes with more realistic shapes (e.g. cylinder, torus),
avoiding the sharp edges and corners of the cubic holes.
Anyhow, they last for a very short period of time 
and therefore do not affect the overall convergence.
On the bottom of the same figure, 
we report the enstrophy difference $E_{2048}(t^*)-E_{512}(t^*)$
for the uncontrolled case (NC) and the one with manifold death (MD). 
At $t^*=2.9$ a large difference (almost one third of the maximum enstrophy value)
can be observed between the uncontrolled cases,
with more enstrophy produced using the coarser grid,
while after $t^*=3.5$ a higher enstrophy production 
can be observed with the finer grid.
This is due to the disruption of the sheet shown in Figure \ref{imm:view} and
without controlled perforation its onset depends on the grid size.
When using the manifold death method, the difference is much smaller
and enstrophy converges. 

To confirm that the cause for the non convergence upon grid refinement of enstrophy is
the breakup of thin sheets in the lighter fluid, we show in Figure \ref{imm:view}
the interface at the dimensionless time $t^*=2.9$ for four cases.
By comparing the two pictures obtained without controlled perforation (top row) the numerical essence of the breakup is clear. 
In fact, at lower resolution (left) the sheet is broken and
many holes and ligaments can be seen,
while at a higher resolution (right) the sheet is well resolved 
since its thickness is larger than the grid size. 
On the contrary, when the manifold death algorithm is used (bottom row),
at the same time the artificial holes have already destroyed 
the sheet in both cases and
the breakup becomes grid independent.
A close-up on the expansion of the holes created
by the manifold death algorithm
and the subsequent ligament formation is shown in Figure \ref{imm:zoom}.
\begin{figure}[htb!]
 \centering
  \includegraphics[width=0.49\textwidth]{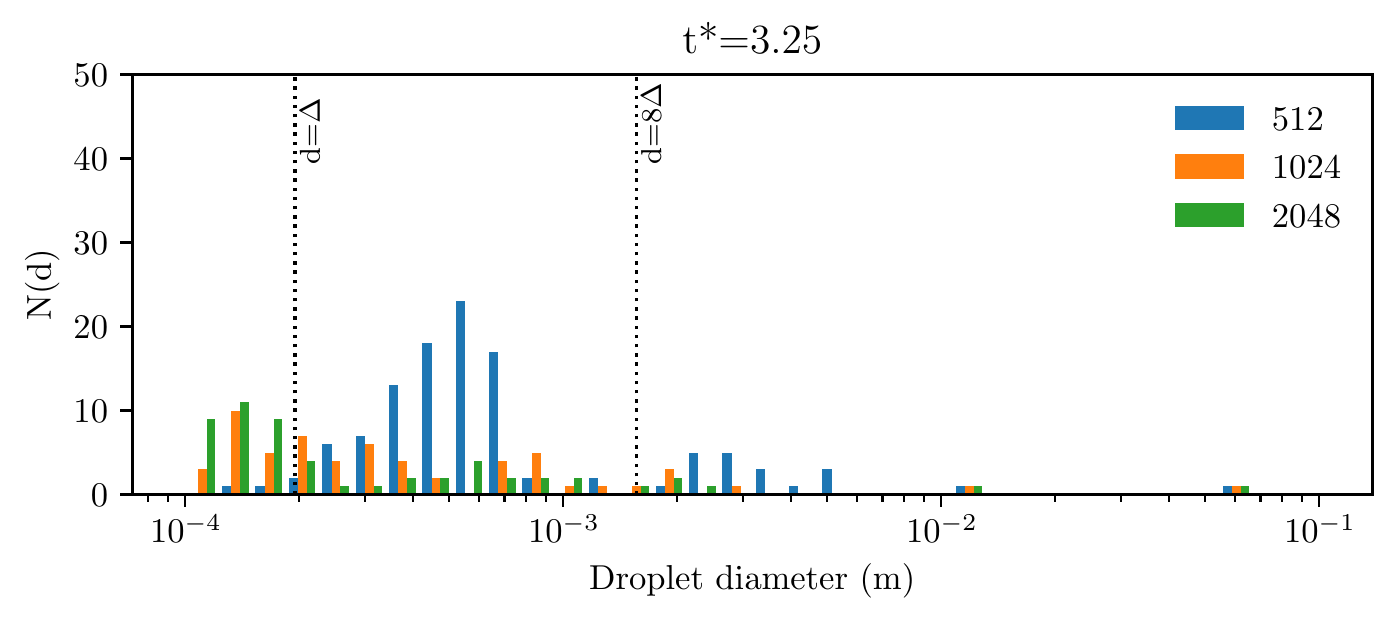}
  \includegraphics[width=0.49\textwidth]{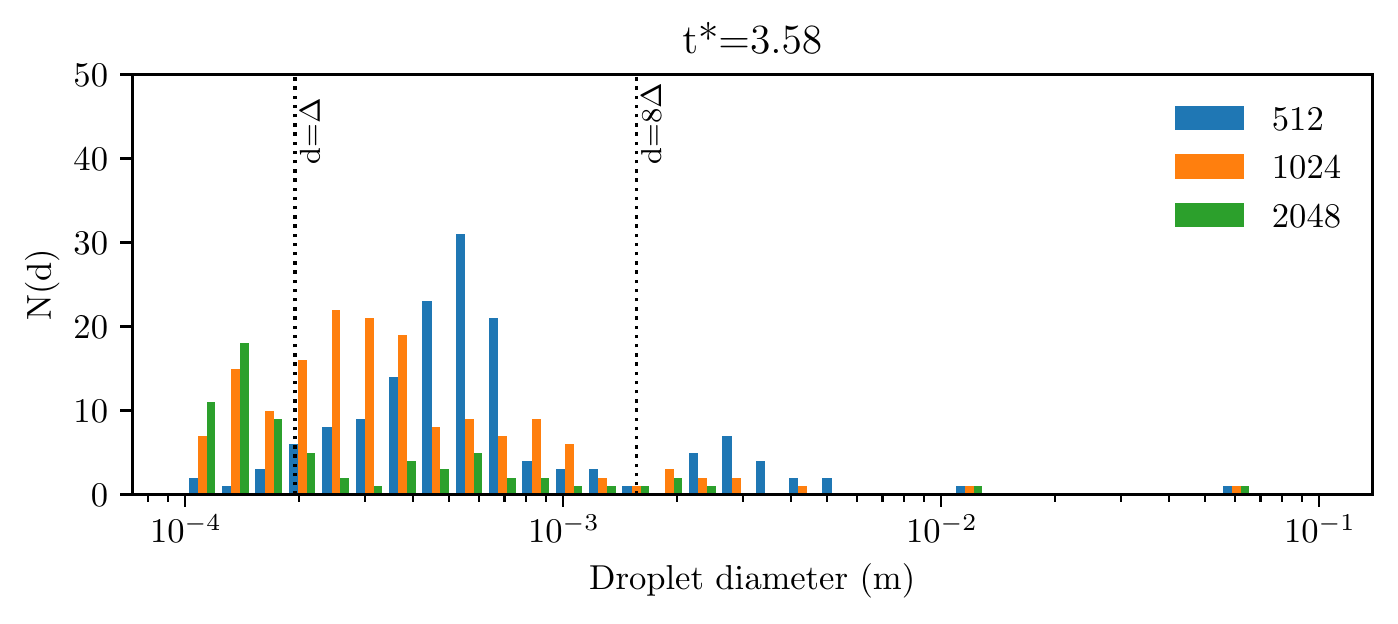}
  \includegraphics[width=0.49\textwidth]{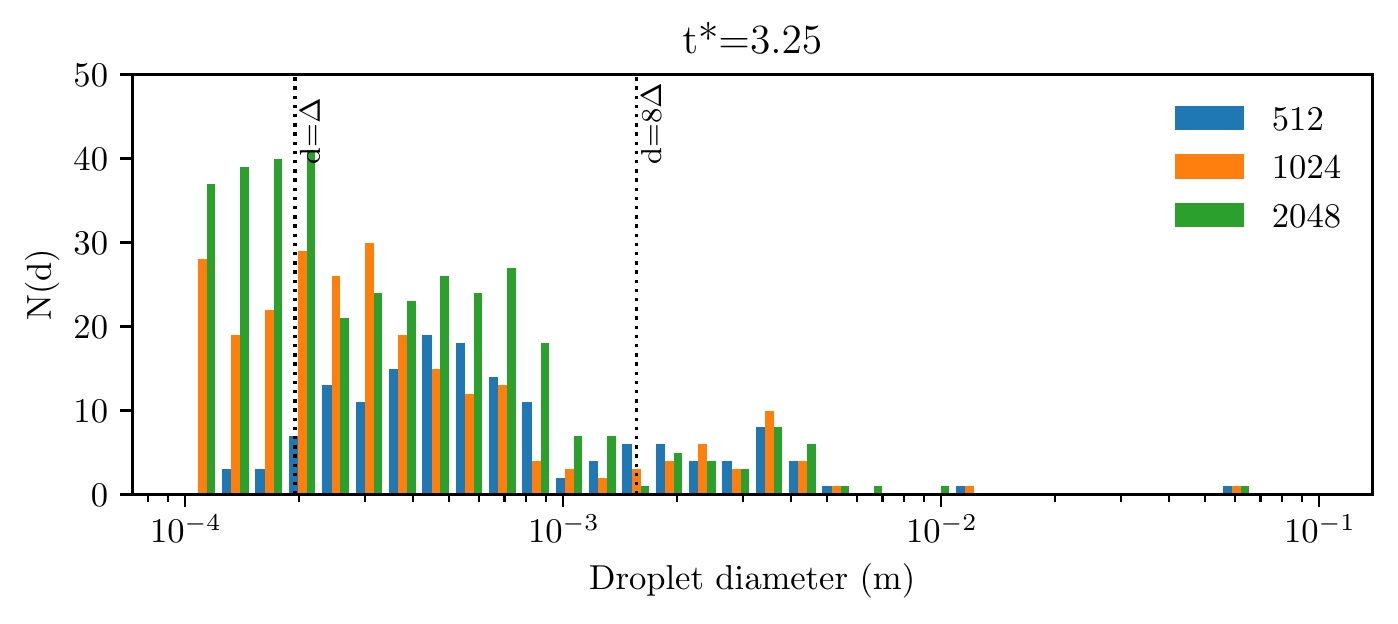}
  \includegraphics[width=0.49\textwidth]{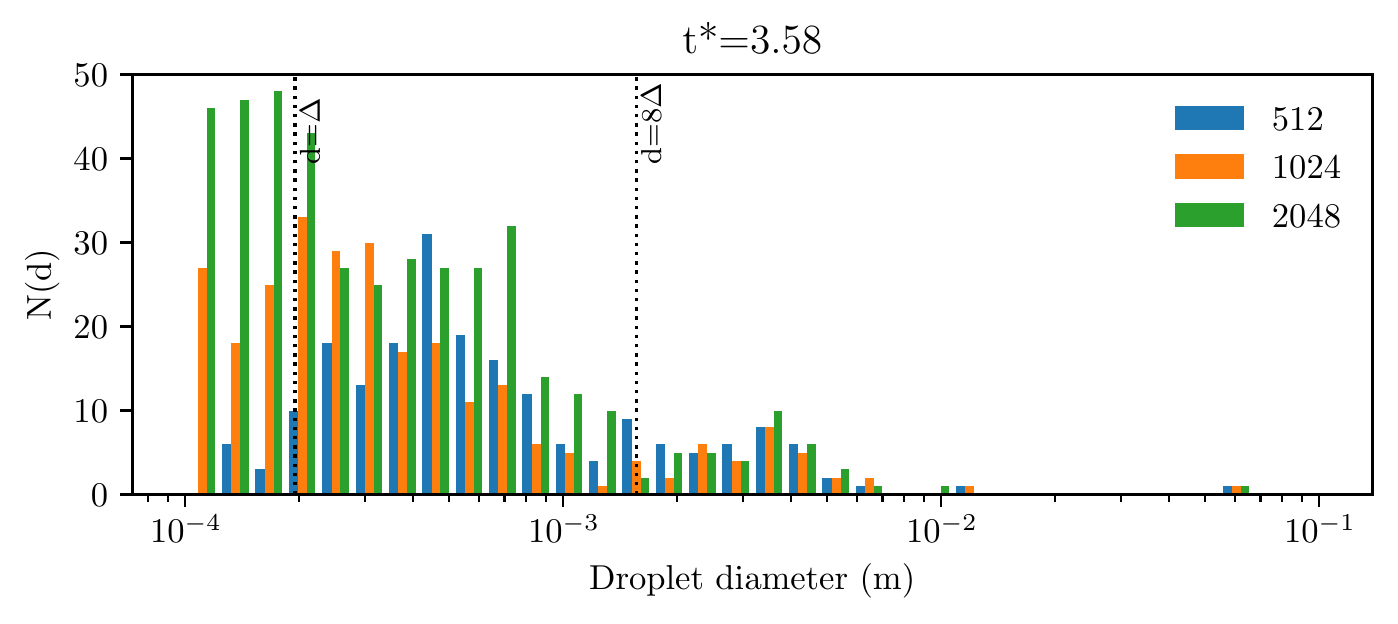}
  \caption{Droplet diameter frequency N(d) without (top) and with (bottom)
  controlled perforation at $t^*=3.25$ and $t^*=3.58$. 
  The droplet diameters are in meters and the box size is $H = 0.1$ m.}
  \label{imm:dropSize}
\end{figure}

In Figure \ref{imm:mass}, the volume deviation of fluid 1
(lighter fluid) is plotted as a function of the dimensionless time.
A staircase profile is obtained for the cases with manifold death, while
the mass is conserved almost exactly in the without controlled perforation. As observed for the enstrophy, 
the frequency of the steps is clearly $1/t^*_h$ and their height is proportional
to the mass lost (up to $h_c^3$ for every hole created). 
The mass lost due to the manifold death algorithm is smaller than 1.3\%
and does almost not depend on the grid used for the simulation.
In future works we plan to improve the method to
redistribute the mass removed from the holes to the surrounding rim.

Finally, we focus on the droplet size distribution and in particular
on the number of larger droplets, since they can be expected to result 
from both physically dominated breakup and manifold death of thin structures.
In Figure \ref{imm:dropSize} we compare the droplet diameter frequency N(d) obtained in the standard case (top) 
and with the use of manifold death (bottom)
as a function of the droplet diameter.
In the former, we recall that at $t^*=3.25$ the sheet is well resolved on the two finer grids, while on the coarser one it has been destroyed by the numerical breakup.
As a consequence, more droplets have been produced using the $512^3$ resolution.
At $t^*=3.58$, the sheet no longer exists also with the $1024^3$ grid.
The two larger droplets ($d>10^{-2}$ m) have been correctly resolved on both grids,
while the frequency of the smaller ones shows similar profiles, with a shift towards
the left (smaller droplets) when the finer mesh is used. 
This confirms the numerical nature of the breakup.
With controlled perforation, the number of trustworthy droplets
on the right of the dashed line ($d>8\Delta_{512}$) is similar
for the three grids and at both times. 
Moreover, when the mesh size decreases, more small droplets 
(in the untrustworthy region $d<\Delta_{512}$) are again collected.
\section{Conclusions}
We have described a new method to detect thin structures and to create holes
depending only on the value of the critical thickness $h_c$.
We have performed a simple single vortex test to prove the ability
to locate thin ligaments of thickness $h_c$ independently on the grid resolution.
The axisymmetric secondary atomization and the three-dimensional phase inversion tests
show the improvements in terms of grid convergence
that the method can provide to the study of complex multiphase simulations.
The findings of this study suggest that using an algorithm to
induce in a controlled way the breakup of thin structures
is necessary to obtain grid independence of the droplet size distribution
and second order moments such as enstrophy.
Future work will concentrate on improving the numerical hole formation,
for example by re-distributing the mass near the holes
rather than removing it, to better conserve mass and energy. 
\section*{Acknowledgements} 
The authors benefited from the ERC grant TRUFLOW. 
The project also benefited from the PRACE grant TRUFLOW number 2020225418
for a large number of CPU hours in 2021.
We are grateful for access to the computational facilities TGCC
granted by GENCI under project number A0092B07760. 
We thank the support of the CSCS Supercomputer Centre 
through the grant s1136, involving the time in the Piz Daint supercomputer.

\bibliography{manifold}

\end{document}